\documentclass[twocolumn]{aastex63}

\usepackage{amsmath}
\usepackage{mathtools}
\usepackage{tabularx}
\usepackage[hyphens]{url}
\usepackage{graphicx}
\usepackage{subfigure}
\usepackage[figuresright]{rotating}
\usepackage{booktabs}
\usepackage{multirow}
\newcommand{\erf}[1]{\operatorname{erf}\left(#1\right)}

\received{XXX}
\revised{XXX}
\accepted{XXX}
\submitjournal{ApJ}

\shorttitle{MaDCoWS DR2}
\shortauthors{Thongkham et al.}

\graphicspath{{./}{figures/}}

\begin{document}

\title{The Massive and Distant Clusters of WISE Survey 2: Second Data Release}

\author[0000-0001-7027-2202]{Khunanon Thongkham}
\affiliation{Department of Astronomy, University of Florida, 211 Bryant Space Center, Gainesville, FL 32611, USA}

\author[0000-0002-0933-8601]{Anthony H. Gonzalez}
\affiliation{Department of Astronomy, University of Florida, 211 Bryant Space Center, Gainesville, FL 32611, USA}
 
\author[0000-0002-4208-798X]{Mark Brodwin}
\affiliation{Department of Physics and Astronomy, University of Missouri, 5110 Rockhill Road, Kansas City, MO 64110, USA}
 
\author[0000-0003-3428-1106]{Ariane Trudeau}
\affiliation{Department of Astronomy, University of Florida, 211 Bryant Space Center, Gainesville, FL 32611, USA}

\author{Peter Eisenhardt}	
\affiliation{Jet Propulsion Laboratory, California Institute of Technology, 4800 Oak Grove Dr., Pasadena, CA 91109, USA}

\author[0000-0003-0122-0841]{S. A.\ Stanford}
\affiliation{Department of Physics and Astronomy, University of California, One Shields Avenue, Davis, CA 95616, USA}

\author[0000-0001-9793-5416]{Emily Moravec}
\affiliation{Green Bank Observatory, P.O. Box 2, Green Bank, WV 24944, USA}

\author[0000-0002-7898-7664]{Thomas Connor}
\affiliation{Center for Astrophysics $\vert$\ Harvard\ \&\ Smithsonian, 60 Garden St., Cambridge, MA 02138, USA}
\affiliation{Jet Propulsion Laboratory, California Institute of Technology, 4800 Oak Grove Dr., Pasadena, CA 91109, USA}

\author[0000-0003-2686-9241]{Daniel Stern}
\affiliation{Jet Propulsion Laboratory, California Institute of Technology, 4800 Oak Grove Dr., Pasadena, CA 91109, USA}

\author[0009-0006-4163-620X]{Ryan Spivey}
\affiliation{Department of Astronomy, University of Florida, 211 Bryant Space Center, Gainesville, FL 32611, USA}

\author[0000-0001-7145-549X]{Karolina Garcia}
\affiliation{Department of Astronomy, University of Florida, 211 Bryant Space Center, Gainesville, FL 32611, USA}

\begin{abstract}
We present the second data release of the Massive and Distant Clusters of WISE Survey 2 (MaDCoWS2). We expand from the equatorial first data release to most of the Dark Energy Camera Legacy Survey area, covering a total area of $6498$ deg$^2$. The catalog consists of $133,036$ S/N $\geq5$ galaxy cluster candidates at $0.1\leq z \leq2$, including $6790$ candidates at $z>1.5$. We train a convolutional neural network (CNN) to identify spurious detections, and include CNN-based cluster probabilities in the final catalog. We also compare the MaDCoWS2 sample with literature catalogs in the same area. The larger sample provides robust results that are consistent with our first data release. At S/N $\geq5$, we rediscover $59-91\%$ of clusters in existing catalogs that lie in the unmasked area of MC2. The median positional offsets are under $250$ kpc, and the standard deviation of the redshifts is $0.031(1+z)$. We fit a redshift-dependent power law to the relation between MaDCoWS2 S/N and observables from existing catalogs. Over the redshift ranges where the surveys overlap with MaDCoWS2, the lowest scatter is found between S/N and observables from optical/infrared surveys. We also assess the performance of our method using a mock light cone measuring purity and completeness as a function of cluster mass. The purity is above $90\%$, and we estimate the $50\%$ completeness threshold at a virial mass of log(M/M$_\odot$)$\approx14.3$. The completeness estimate is uncertain due to the small number of massive halos in the light cone, but consistent with the recovery fraction found by comparing to other cluster catalogs. 
\end{abstract}


\section{Introduction} \label{sec:Introduction}

The Massive and Distant Clusters of \textit{WISE} Survey 2, or MaDCoWS2, is a galaxy cluster survey based on galaxy photometric redshifts derived from CatWISE \citep{2020Eisenhardt,2021Marocco} and the Dark Energy Camera Legacy Survey \citep[DECaLS;][]{2019Dey}. With the positions of \textit{WISE} detections as the basis of the survey, MaDCoWS2 uses PZWav \citep{2014Gonzalez,2019Euclid} to find clusters at $0.1\leq z\leq2$ within a region covering more than $50\%$ of the DECaLS area. The goal of the survey is to provide a large catalog of galaxy clusters that extends from low redshift to $z>1.5$ over a very large area.

In the first data release of MaDCoWS2 \citep[DR1;][]{2024Thongkham}, we focused on an $1838$ deg$^2$ equatorial region enclosing the wide survey of the HSC Subaru Strategic Program fields \citep{2018Aihara,2018Aiharab}. The DR1 catalog
provided a catalog of $22,970$ cluster candidates with signal-to-noise ratios (based on Poisson background noise) of S/N$_{\rm P}$ $\geq5$ at $0.1\leq z\leq2$, including $1312$ cluster candidates at $z>1.5$.  

In this paper, we present the second data release of MaDCoWS2 (DR2). DR2 uses most of the DECaLS coverage area, excluding regions near the DECaLS coverage boundaries. The total area of DR2 is $8436$ deg$^2$, with an effective area of $6498$ deg$^2$ after masking. As in DR1 \citep{2024Thongkham}, we cross-match the MaDCoWS2 catalog with existing cluster catalogs from the literature. For DR2, the external catalogs in this comparison include the ACT Sunyaev–Zel’dovich effect Data Release 5 cluster survey \citep{2021Hilton}, the South Pole Telescope-Sunyarv-Zel'dovich catalog \citep[SPT;][]{2015Bleem,2019Bocquet,2020Huang,2020Bleem,2024Bleem}, the \textit{Planck} 2$\text{nd}$ Sunyaev-Zeldovich Source Catalog \citep[PSZ2;][]{2016Planck}, the extended ROentgen Survey with an Imaging Telescope Array (eROSITA) All-Sky Survey  cluster catalog \citep[eRASS;][]{2024Bulbul,2024Kluge}, the red-sequence matched-filter Probabilistic Percolation catalog from DES Science Verification data \citep[redMaPPer;][]{2016Rykoff}, and MaDCoWS \citep{2019Gonzalez}.

A new feature of the DR2 catalog is that we use a Convolutional Neural Network (CNN) to flag detections at $z>1$ that are potentially spurious due to foreground contamination or artifacts from the photometric catalogs. We also assess the purity and completeness of MaDCoWS2 by conducting a search on a mock catalog of galaxies with photometric data quality similar to our search \citep{2023Yung}.

The structure of the paper is as follows. In section \ref{sec:survey overview}, we give a brief overview of the survey. Section \ref{sec:CNN} explains how we use the CNN to identify spurious detections. The basic properties of our catalog are discussed in section \ref{sec:Basic characteristics}, while the characterization of the catalog is in section \ref{sec:survey characterization}. Section \ref{sec:simulation} explores the performance of our method in simulated data, while section \ref{sec:summary} summarizes our results.

We assume a flat $\Lambda$CDM cosmology from \cite{2020Planck} with $\Omega_m =0.315$ and $H_0 = 67.4$ km s$^{-1}$ Mpc$^{-1}$. Magnitudes are in the Vega system unless stated otherwise.

\begin{figure*}[htbp]
\centering
\subfigure{\includegraphics[width=0.9\textwidth]{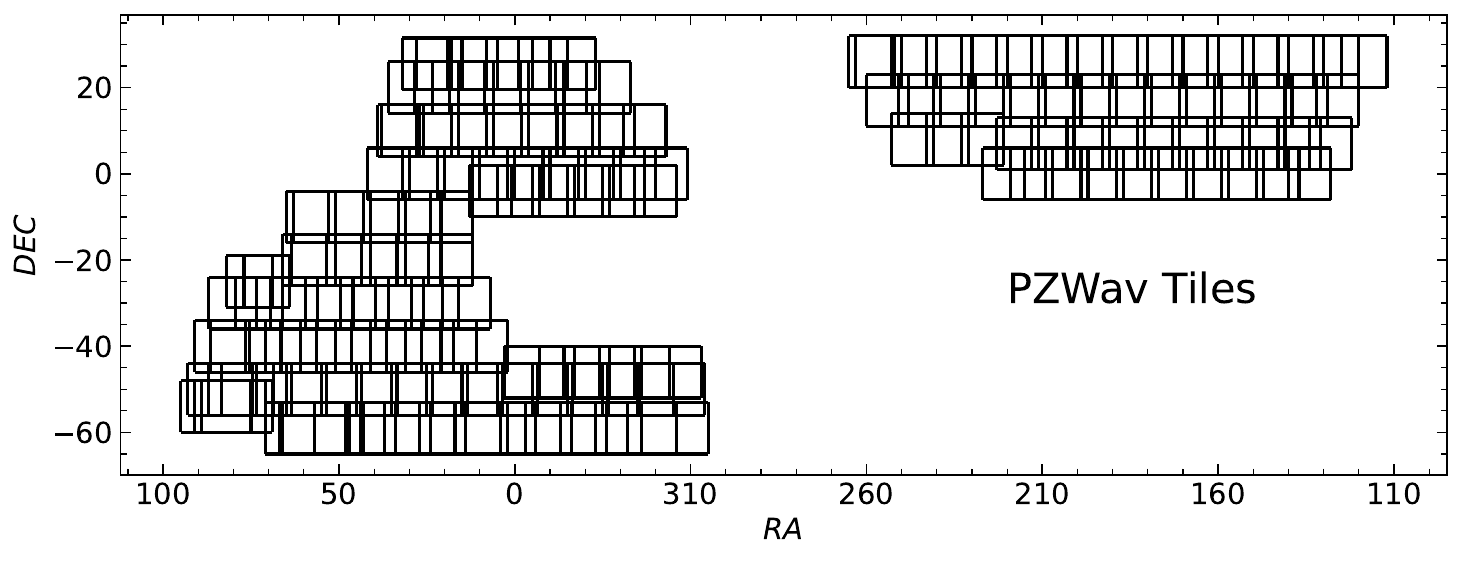}}
\subfigure{\includegraphics[width=0.9\textwidth]{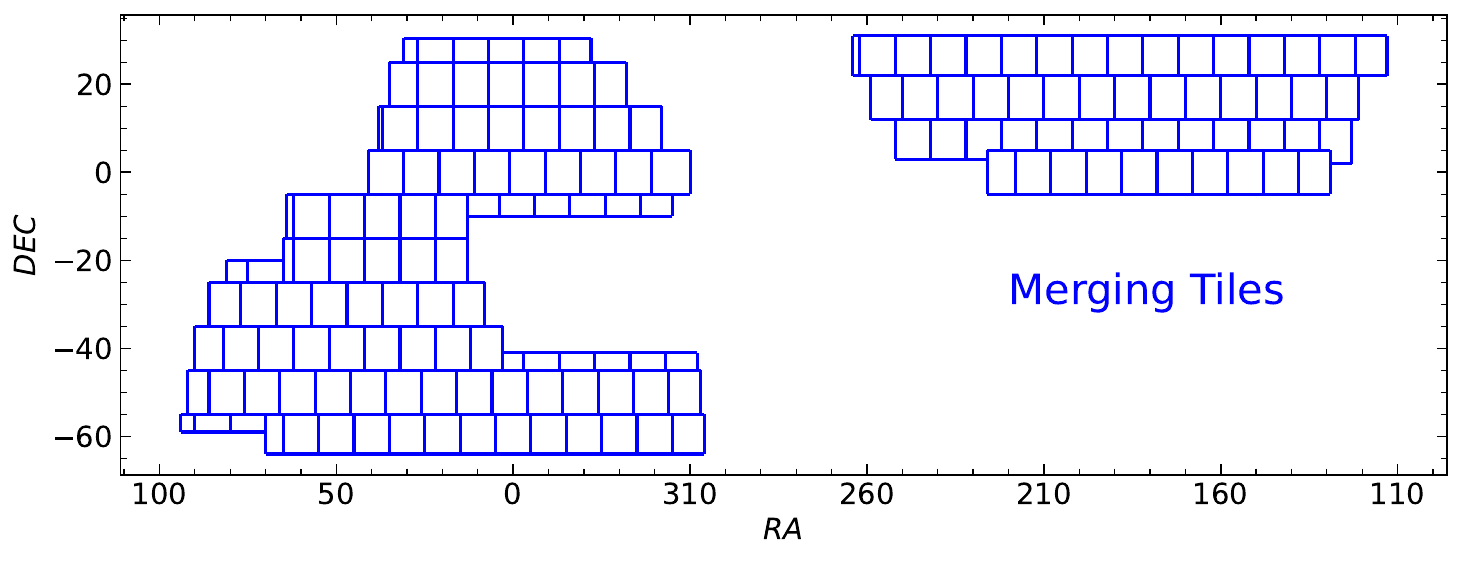}}
\caption{The tiling for this work. We run PZWav on each of the black rectangles independently, and merge the results using the tiling from the blue rectangles.} 
\label{fig:tiles}
\end{figure*} 

\section{Survey Overview} \label{sec:survey overview}
This section provides a brief overview and the update of the survey. A full description of data preparation and the cluster finding process of MaDCoWS2 is available in \cite{2024Thongkham}. 
\subsection{Data} \label{sec:data}
The data preparation of the full release of MaDCoWS2 is the same as in DR1. Our cluster finding algorithm uses a full probability density function (PDF) of the redshift of each galaxy as input. To generate the PDF, we combine the $g$, $r$, and $z$ band data from the DESI Legacy imaging survey DR9 (LS) with the W1 and W2 data from CatWISE2020. Specifically, we use the DECaLS data from LS. We match objects from DECaLS to CatWISE2020 using the positions from CatWISE2020 as a basis. The probability that a match is real is calculated based on the ratio of the separation distribution of real and random matches where random matches are created by shifting CatWISE2020 sources by $5^{\prime}$ (see \S 3.1 of \citealp{2024Thongkham} for more detail). Only matches that have more than $50\%$ probability of being real are used in our search. 

We generate photometric redshifts for each galaxy by computing the $\chi^2$ of the combined photometry with respect to a subset of the empirical templates in \citet{2007Polletta} and the elliptical template of \citet{1980CWW}. The templates are extended to the IR and UV using a stellar population synthesis model (SPS) with an exponentially decaying star formation rate with $\tau = 1$ Gyr, based on \citet{2003Bruzual} models. We project $\chi^2$ along the redshift dimension to create a full PDF for each galaxy. The full details on the photometric redshifts used in MaDCoWS2 will be presented in a forthcoming paper.

We remove potentially problematic objects from our input catalog using flags from CatWISE2020 and the LS galaxy model.  Stars are also removed based on $r-z$ and $z-W1$ colors. We refer readers to \S 3.2 of \cite{2024Thongkham} for more details on the filtering process.

\subsection{Cluster Finding} \label{sec:cluster finding} 
\subsubsection{PZWav} \label{subsec:algorithm}
MaDCoWS2 uses PZWav \citep{2014Gonzalez,2019Euclid} to identify overdensities of galaxies in three dimensions (RA, DEC, and redshift). The algorithm is one of the two cluster detection algorithms of the \textit{Euclid} mission \citep{2019Euclid} and is also used by \citet{2023Werner} for the S-PLUS survey. A forerunner of PZWav was initially developed for the \textit{Spitzer} IRAC Shallow Cluster Survey (ISCS) in \cite{2008Eisenhardt} and the \textit{Spitzer} IRAC Deep Cluster Survey (IDCS) in \cite{2012Stanford}. We refer to \cite{2024Thongkham} for a complete explanation of the use of PZWav in MaDCoWS2, and provide only a brief description here. PZWav takes a galaxy catalog with coordinates and full PDFs of photometric redshifts as input. The algorithm constructs a density cube that is then smoothed by a difference-of-Gaussian wavelet kernel. The density cube is masked by galaxy and star masks to reduce contamination (see \S 5.3 of \citealp{2024Thongkham}). We also mask globular clusters and planetary nebulae larger than 5$^{\prime\prime}$ using the catalog available with LS\footnote{\url{https://portal.nersc.gov/cfs/cosmo/data/legacysurvey/dr10/masking/NGC-star-clusters.fits}}. Galaxy clusters are then detected as the highest peaks in the density cube. Each detection has an S/N where the noise is calculated from a bootstrap noise map assuming either Poisson or Gaussian background noise (S/N$_{\rm P}$ and S/N$_{\rm G}$). The photometric redshift of each candidate is refined using the sigma-clipped median redshifts of galaxies around the detection position.

\subsubsection{Tiling} \label{subsec:tiling}
For DR2, we employ the same tiling scheme that was used in DR1, but expand the search area to the full coverage area of DECaLS. This scheme is shown in Figure \ref{fig:tiles}. Each black tile in the top panel has an approximate area of $140$ deg$^2$. We remove edges and overlapping regions in the black tiles to obtain the blue tiles which we combine to create the final catalog. The catalog is filtered after merging to remove detections around edges or close to masks as described in \S 5.3 of \cite{2024Thongkham}.

We applied no hard restriction on Galactic latitude during this search. Thus, approximately 1$\%$ of the cluster candidates lie at $|b|<25^\circ$. Candidates at such low Galactic latitude should be considered with caution, as the higher stellar density near the Galactic plane may impact the S/N of detections due to cluster members being blended with foreground stars, or may lead to elevated contamination from artifacts. Thus, for statistical analyses, we advise the use of clusters only at higher Galactic latitude. The total effective area of the search at $|b|>25^\circ$ is $6433$ deg$^2$.

\section{Convolutional Neural Network} \label{sec:CNN}
Occasionally, fragmented galaxies and bright star artifacts in CatWISE2020 are not removed by algorithms used for masking and rejection. Because the galaxy fragments and star artifacts   are typically matched to incorrect or faint optical counterparts, they have the potential to result in spurious cluster detections at $z\geq1$. We employ a convolutional neural network (CNN) to identify these remaining spurious detections. We fine tune a pre-trained model called ResNet50 \citep{2015He} from \textsc{fastai} \citep{2020Howard} using images of genuine cluster candidates, spurious detections from large galaxies, and spurious detections from bright stars. The large galaxies have half-light radii $R_{e}>3\farcs5$. All input images are visually inspected to verify their classifications. The training data for the CNN consist of 1136, 84, and 220 images of clean candidates, candidates with large galaxies, and candidates with bright stars, respectively. To augment the number of the spurious candidates, we rotate and flip the images of the spurious candidates resulting in $336$ and $880$ training images for detections with big galaxies and bright stars, respectively. We use $70\%$ of the data set for training and $30\%$ for validation.

\begin{figure}[htbp]
\centering
\includegraphics[width=0.9\columnwidth]{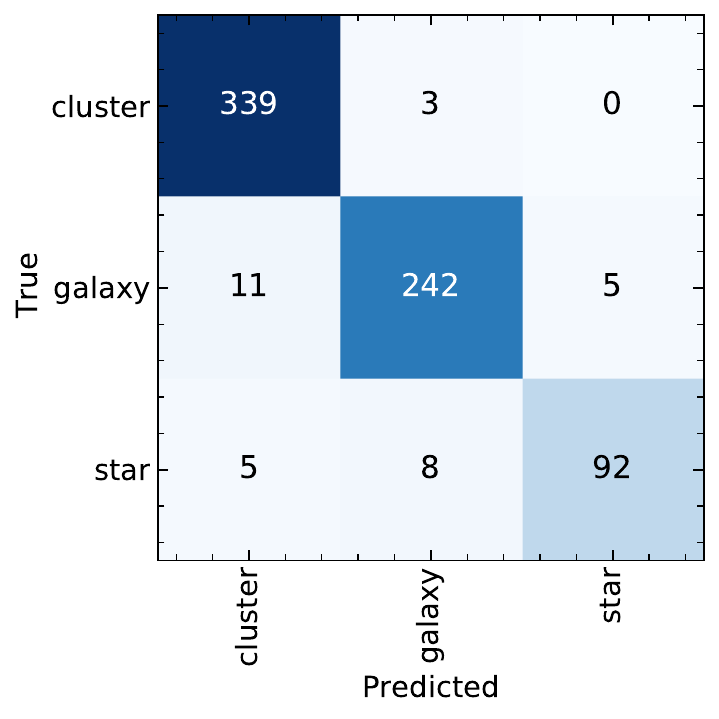}
\caption{Confusion matrix of genuine cluster detection, spurious detection from large galaxies, and spurious detection from bright stars. The majority of the validation data lie on the diagonal elements of the matrix showing the capability of the CNN to identify spurious detections in our catalog.}
\label{fig:confusion matrix}
\end{figure}

We fine tune our CNN model for nine epochs to reach $95\%$ accuracy. Figure \ref{fig:confusion matrix} shows the confusion matrix for the model displaying the capability of the CNN to identify each type of spurious detection. We include the probability that a cluster detection is not spurious as a column named P$_{\rm CNN}$ in our catalog for cluster candidates at $z\geq1$. We only use cluster candidates with P$_{\rm CNN}>0.5$ during comparison with external catalogs (\S \ref{sec:survey characterization}). We recommend readers to use only cluster candidates with P$_{\rm CNN}>0.5$ if purity is of concern.

\section{The Catalog} \label{sec:Basic characteristics}
We present a catalog of $133,036$ cluster candidates with S/N$_{\rm P}$ $\geq5$ in the MaDCoWS2 full data release. The catalog covers redshifts ranging from $0.1$ to $2$, including $29,764$ candidates at $z\geq1$ and $6790$ candidates at $z>1.5$. The DR2 catalog is $5.79$ times larger than DR1, enabling more definitive assessment of the properties of the catalog. The catalog consists of name, RA, DEC, photometric redshift (and its error) and the S/N based upon Gaussian (S/N$_{\rm G}$) and Poisson (S/N$_{\rm P}$) noise statistics. In addition, we include the names in the literature of each cluster candidate (and numbers assigned to the referred works) and spectroscopic redshifts derived from external catalogs accompanied by their references. Lastly, we provide the name of any foreground cluster detections in the column $\rm Name_{fg}$ (see \S \ref{subsec:algorithm} of MaDCoWS2 DR1) and the CNN probability that a detection $at z>1$  is a genuine cluster P$_{\rm CNN}$ (see \S \ref{sec:CNN}). A foreground detection is defined as a cluster candidate within a projected $500$ kpc of a more distant candidate, requiring a redshift at least $0.12$ (two bins) lower than that of the more distant candidate. The full table of the DR2 catalog is available electronically and included with this paper as a machine-readable table. The 10 highest S/N$_{\rm P}$ MaDCoWS2 clusters are presented in Table \ref{tab:sample2}. In the analysis below, we use $130,723$ cluster candidates which consist of $27,451$ at $z\geq1$ with P$_{\rm CNN}>0.5$.

\begin{figure*}[htbp!]
\centering
\subfigure{\includegraphics[width=0.9\textwidth]{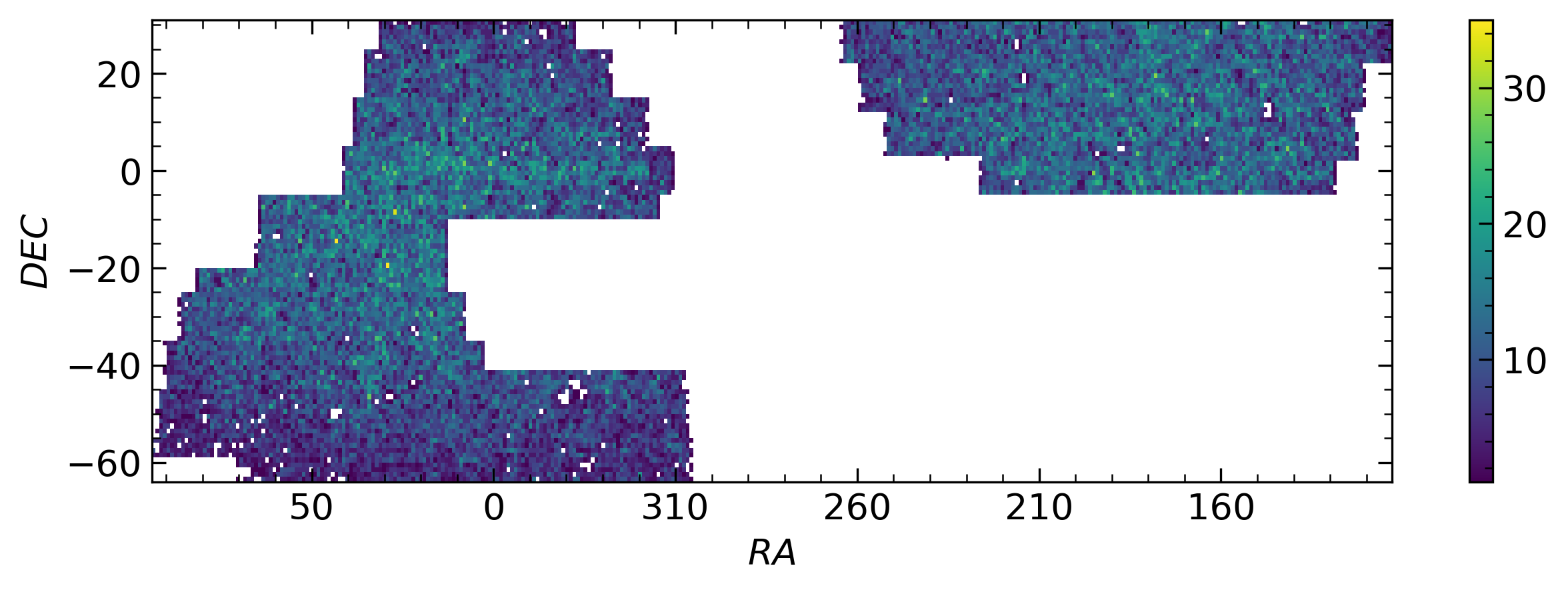}}
\caption{Sky distribution of MaDCoWS2 cluster candidates. The colors indicate the number of cluster candidates in each bin of the two-dimensional histogram. Spatial bins span 1 degree in both right ascension and declination and are not adjusted for the decrease in area with increasing absolute declination.} 
\label{fig:spatdist}
\end{figure*}

We display the spatial distribution of the MaDCoWS2 full data release in Figure \ref{fig:spatdist}. The area is $6498$ deg$^2$ after masking ($8436$ deg$^2$ including masked area). We note that the MaDCoWS2 area is smaller than the total area of LS because we avoid searching for clusters close to the edges of LS where the optical data are not complete and the LS depth varies significantly.

\begin{rotatetable*}
\begin{deluxetable*}{llrcccccclccc}
\centerwidetable
\tabletypesize{\footnotesize}
\tablecaption{MaDCoWS2 galaxy cluster candidates, in order of descending S/N$_{\rm P}$. \label{tab:sample2}}
\tablehead{\colhead{Name} & \colhead{RA} & \colhead{DEC} & \colhead{$z_{\rm phot}$} & \colhead{$\epsilon_{z_{\rm phot}}$} & \colhead{S/N$_{\rm P}$} & \colhead{S/N$_G$} & \colhead{Name$_{\rm fg}$} & \colhead{P$_{\rm CNN}$} &\colhead{Literature Name} & \colhead{LitRef} & \colhead{zspec} & \colhead{zspecRef} \\ 
\colhead{} & \colhead{Deg} & \colhead{Deg} & \colhead{} & \colhead{} & \colhead{} & \colhead{} & \colhead{} & \colhead{} & \colhead{} & \colhead{} & \colhead{} & \colhead{}} 
\startdata
MOO2 J21503-08473 &  327.588 & -8.790 & 0.380 & 0.037 & 21.5 & 33.7 & \nodata & \nodata & ACT-CL J2150.3-0847 & [2,16,20,24] & 0.395 & 2 \\
MOO2 J22419+17326 & 340.483 &  17.545 & 0.310 & 0.035 & 21.3 & 32.8 & \nodata & \nodata & ACT-CL J2241.9+1732 & [2,4,16,20,24] & 0.313 & 2 \\
MOO2 J02031-20172 &  30.789 & -20.288 & 0.420 & 0.038 & 20.3 & 31.4 & \nodata & \nodata & ACT-CL J0203.1-2017 & [2,8,23,24] & \nodata & \nodata \\
MOO2 J02199+01304 &  34.976 &  1.507 & 0.350 & 0.036 & 20.2 & 30.6 & \nodata & \nodata & ACT-CL J0219.9+0130 & [2,16,18,20,23,24] & 0.365 & 2\\
MOO2 J22117-03491 &  332.945 &  -3.819 & 0.410 & 0.038 & 20.1 & 31.0 & \nodata & \nodata & ACT-CL J2211.7-0349 & [2,4,16,20,24] & 0.428 & 20\\
MOO2 J23083-02114 &  347.084 &  -2.190 & 0.310 & 0.035 & 20.1 & 30.5 & \nodata & \nodata & ACT-CL J2308.3-0211 & [2,4,16,20,23,24] & 0.289 & 2\\
MOO2 J09498+17069 &  147.470 & 17.115 & 0.370 & 0.037 & 19.7 & 29.8 & \nodata & \nodata & ACT-CL J0949.8+1707 & [2,4,6,8,16,20,24] & 0.388 & 2\\
MOO2 J02398-01345 &  39.972 &  -1.576 & 0.350 & 0.036 & 19.7 & 29.6 & \nodata & \nodata & Abell  370 & [1,2,4,16,20,23,24] & 0.373 & 1\\
MOO2 J04112-48194 &  62.811 & -48.324 & 0.380 & 0.037 & 19.5 & 28.7 & \nodata & \nodata & ACT-CL J0411.2-4819 & [2,3,4,8,23,24] & 0.424 & 2\\
MOO2 J00407-44082 &  10.196 &  -44.137 & 0.340 & 0.036 & 18.8 & 27.4 & \nodata & \nodata & ACT-CL J0040.8-4407 & [2,3,4,6,8,23,24] & 0.350 & 2 \\
\enddata
\tablecomments{Example lines from the full MaDCoWS2 catalog, showing the 10 highest S/N$_{\rm P}$ candidates. The full table is provided electronically in machine-readable form. Reference: [1] \cite{1989Abell}, [2] \cite{2021Hilton}, [3] \cite{2019Bocquet}, [4] \cite{2016Planck}, [5] \cite{1990Gioia}, [6] \cite{2001Ebeling}, [7] \cite{2005Gladders}, [8] \cite{2024Bulbul}, [9] \cite{2021Balogh}, [10]\cite{2012Muzzin,2017Balogh}, [11] \cite{2009Andreon}, [12] \cite{2010Papovich}, [13] \cite{2018Adami}, [14] \cite{2022Liu}, [15] \cite{2012Mehrtens}, [16] \cite{2016Rykoff}, [17] \cite{2019Gonzalez}, [18] \cite{2018Oguri}, [19] \cite{2017Radovich}, [20] \cite{2012Wen}, [21] \cite{2015Wen}, [22] \cite{2021Wen}, [23] \cite{2022Wen}, [24] \cite{2024Wen}.}
\end{deluxetable*}
\end{rotatetable*}

\begin{figure}[htbp]
\centering
\includegraphics[width=0.9\columnwidth]{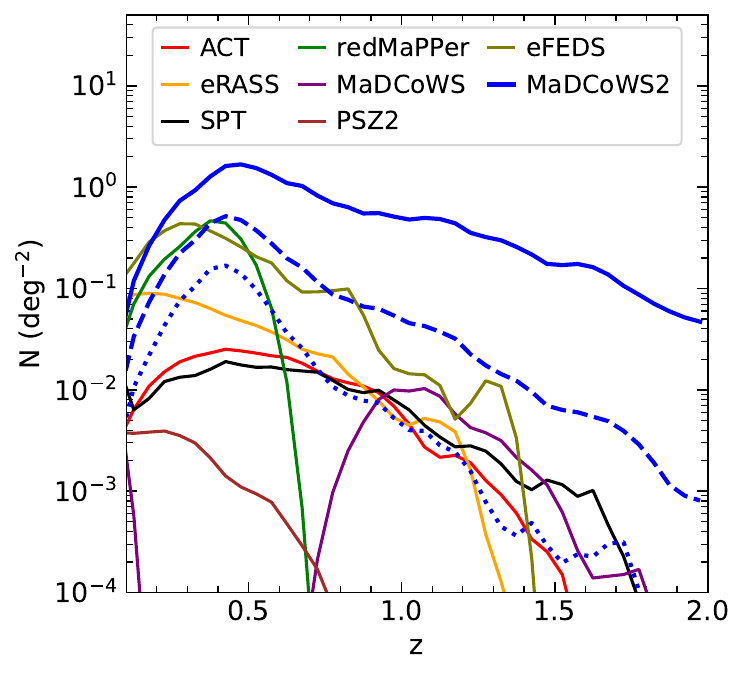}
\caption{Photometric redshift distribution of MaDCoWS2 and external cluster catalogs normalized by their coverage areas. The solid, dashed, and dotted lines correspond to S/N$_{\rm P}$$\geq5$, $7$, and $9$, respectively, for MaDCoWS2. The distributions are smoothed by a Gaussian filter with a standard deviation of $0.0375$. MaDCoWS2 provides significantly more cluster candidates at $z>1.5$ compared to other catalogs.} 
\label{fig:zdist2}
\end{figure}

The estimated redshift distributions normalized by coverage areas of MaDCoWS2 and external catalogs are shown in Figure \ref{fig:zdist2}\footnote{The normalization for MaDCoWS2 in DR2 uses the effective search area. This normalization is different from the normalization in DR1 which is based on the search area including the masked regions.}. Similar to DR1, the redshift distribution of MaDCoWS2 peaks at $z \sim 0.5$. The limited comoving volume at low redshift reduces the number of detections per unit area. At high redshift, the smaller number of detections results from a combination of real cluster mass evolution and the diminishing S/N$_{\rm P}$ at fixed cluster mass. We include the eFEDS clusters \citep{2022Liu} in Figure 4 because the eRASS survey (see \S 5.1.4, \citealp{2024Merloni}) is shallower than eFEDS despite both being selected using eROSITA. As shown in DR1, MaDCoWS2 has comparable depth to eFEDS at $z<0.5$.

\begin{figure}[htbp]
\centering
\includegraphics[width=0.9\columnwidth]{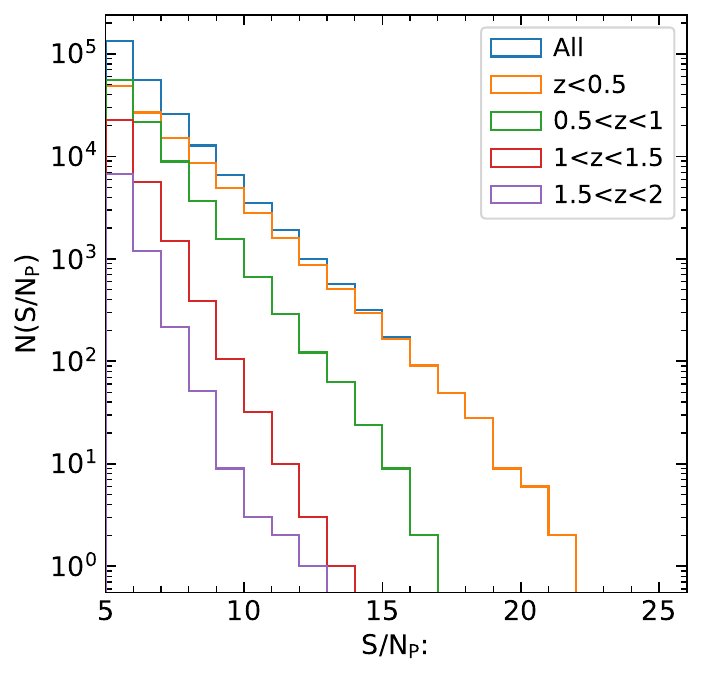}
\caption{Cumulative S/N$_{\rm P}$ distribution for MaDCoWS2 cluster candidates as a function of redshift. More than 200 cluster candidates with S/N$>7$ exist at $z>1.5$} 
\label{fig:SNRdist}
\end{figure}

The distribution of the S/N$_{\rm P}$ of MaDCoWS2 cluster candidates is shown in Figure \ref{fig:SNRdist} as a cumulative histogram. MaDCoWS2 finds $204$ S/N$>7$ cluster candidates at $z>1.5$. The S/N decreases as the redshift increases reflecting that fact that the number of bright cluster galaxies decreases at high redshift.

\begin{figure*}[htbp]
\begin{center}
\begin{tabularx}{\textwidth}{ccc}
\includegraphics[width=0.3\textwidth]{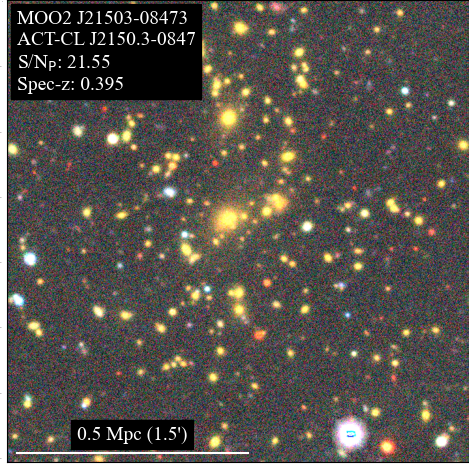} & 
\includegraphics[width=0.3\textwidth]{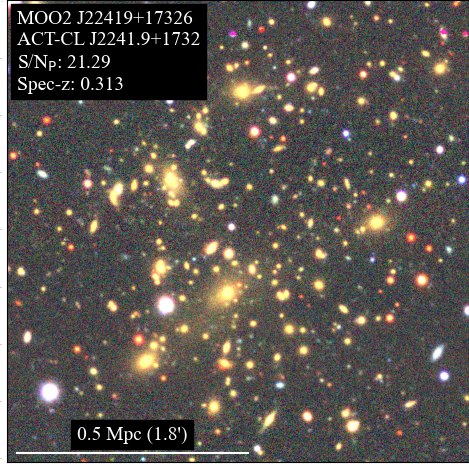} &
\includegraphics[width=0.3\textwidth]{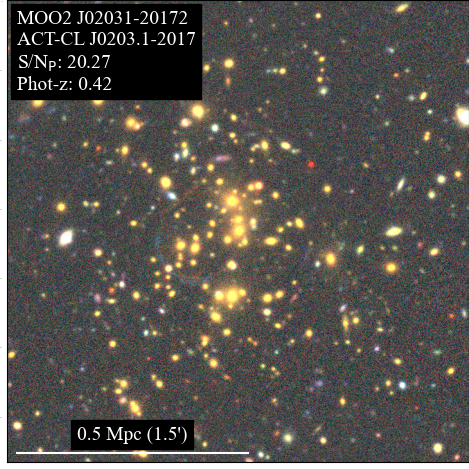}
\end{tabularx}
\begin{tabularx}{\textwidth}{ccc}
\includegraphics[width=0.3\textwidth]{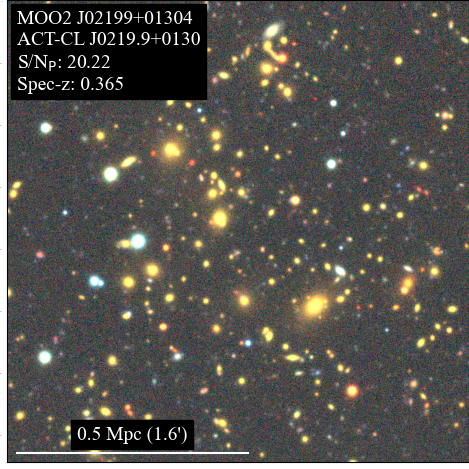} & 
\includegraphics[width=0.3\textwidth]{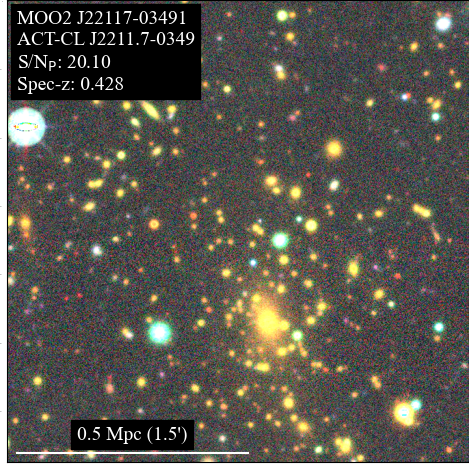} &
\includegraphics[width=0.3\textwidth]{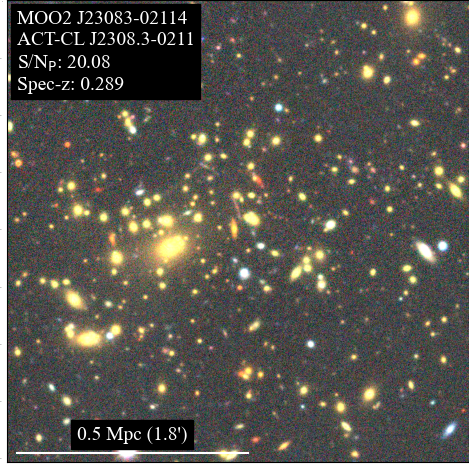}
\end{tabularx}
\begin{tabularx}{\textwidth}{ccc}
\includegraphics[width=0.3\textwidth]{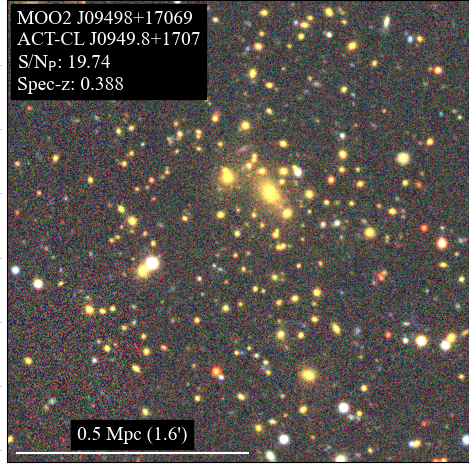} & 
\includegraphics[width=0.3\textwidth]{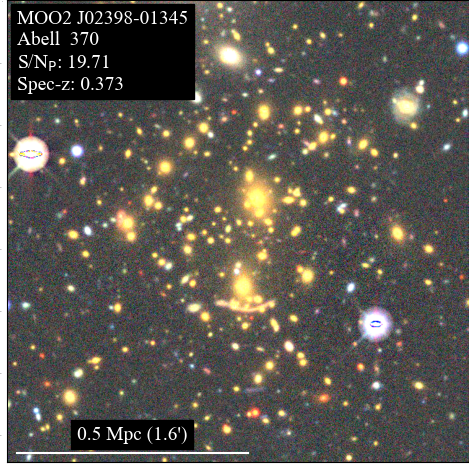} &
\includegraphics[width=0.3\textwidth]{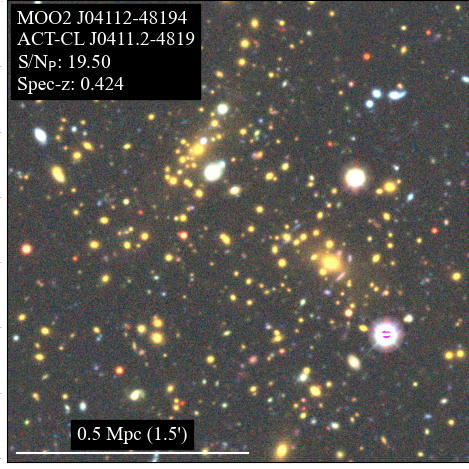}
\end{tabularx}
\begin{tabularx}{\textwidth}{ccc}
\includegraphics[width=0.3\textwidth]{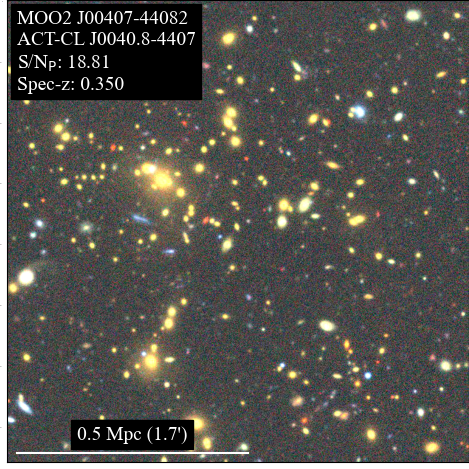} & 
\includegraphics[width=0.3\textwidth]{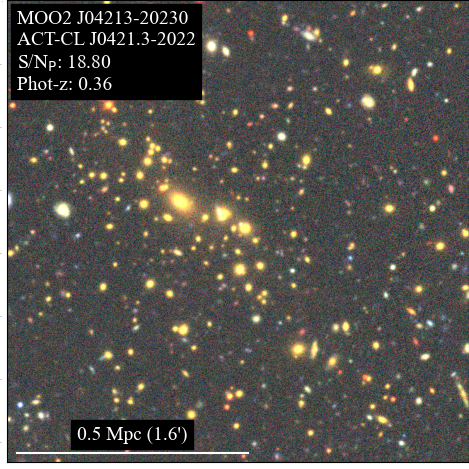} &
\includegraphics[width=0.3\textwidth]{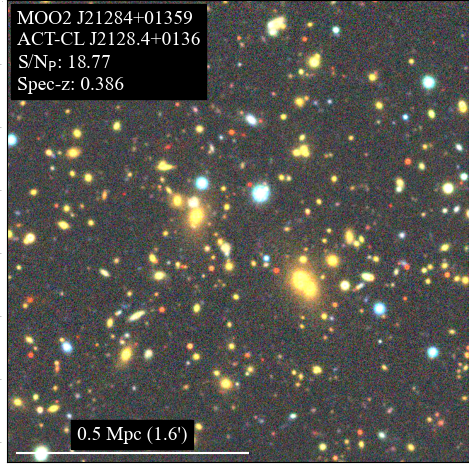}
\end{tabularx}
\caption{\label{fig:pic0} LS DR9 \textit{grz} images of the twelve highest S/N$_{\rm P}$ detections in the MaDCoWS2 catalog. The size of the images is $1 \text{~Mpc} \times 1 \text{~Mpc}$. In each panel, we provide the name of the cluster candidate, the name of its ACT counterpart, our S/N$_{\rm P}$, and our photometric redshift (or spectroscopic redshift if available from an external catalog.)}
\end{center}
\end{figure*}

Figure \ref{fig:pic0} shows DECaLS $g$, $r$, and $z$ images of the twelve MaDCoWS2 cluster candidates with the highest S/N$_{\rm P}$ at $0.1\leq z\leq2$. All twelve candidates are confirmed by ACT DR5, including ten with spectroscopic redshifts. This illustrates the ability of MaDCoWS2 to detect bona fide clusters at low redshift.

\begin{figure*}[htbp]
\begin{center}
\begin{tabularx}{\textwidth}{ccc}
\includegraphics[width=0.31\textwidth]{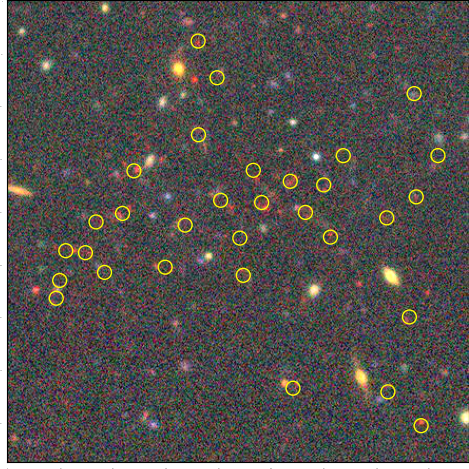} & 
\includegraphics[width=0.31\textwidth]{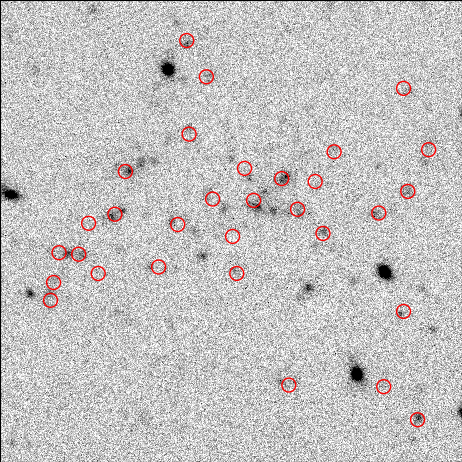} &
\includegraphics[width=0.31\textwidth]{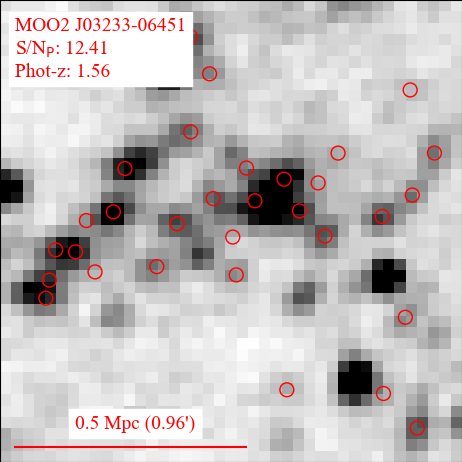}
\end{tabularx}
\begin{tabularx}{\textwidth}{ccc}
\includegraphics[width=0.31\textwidth]{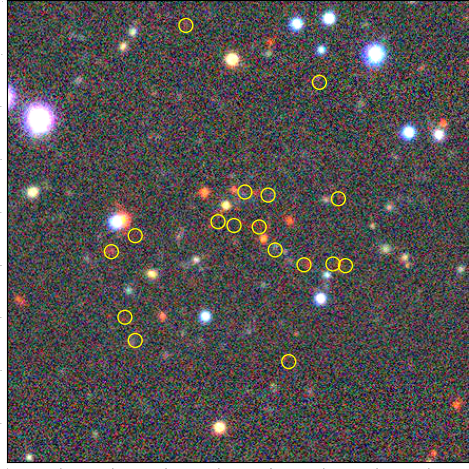} & 
\includegraphics[width=0.31\textwidth]{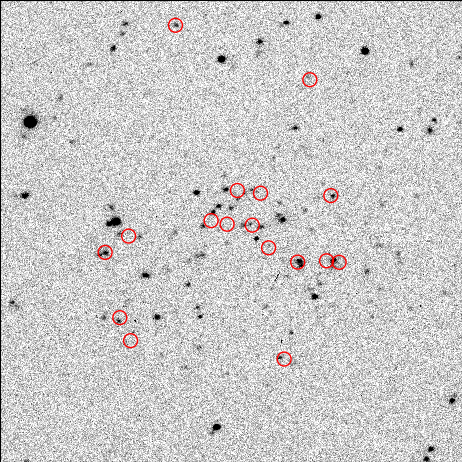} &
\includegraphics[width=0.31\textwidth]{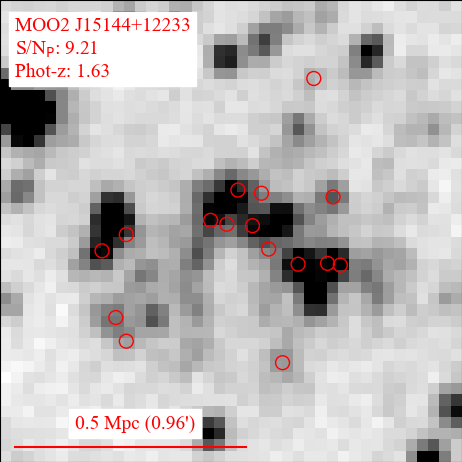}
\end{tabularx}
\begin{tabularx}{\textwidth}{ccc}
\includegraphics[width=0.31\textwidth]{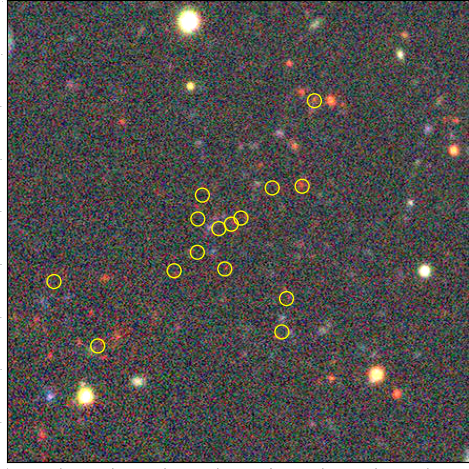} & 
\includegraphics[width=0.31\textwidth]{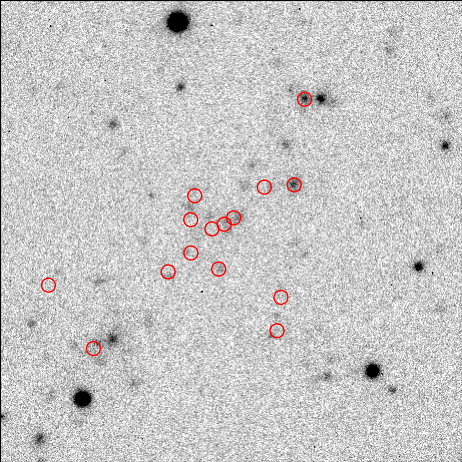} &
\includegraphics[width=0.31\textwidth]{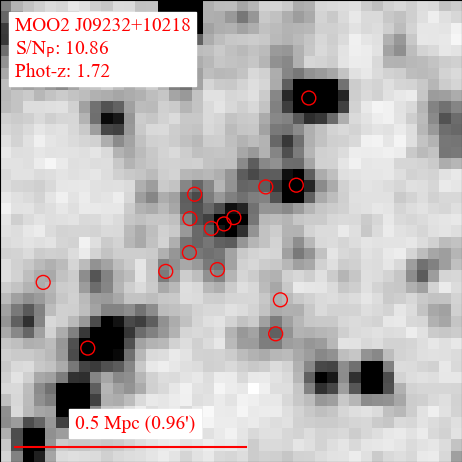}
\end{tabularx}
\begin{tabularx}{\textwidth}{ccc}
\includegraphics[width=0.31\textwidth]{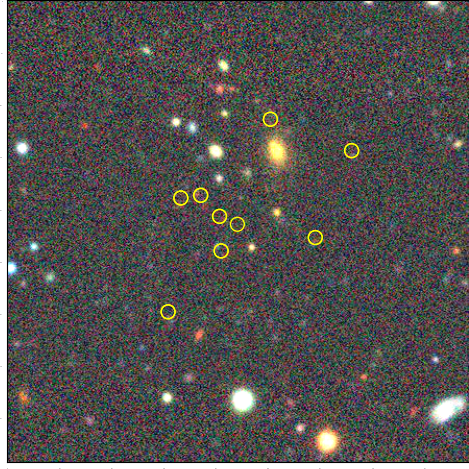} & 
\includegraphics[width=0.31\textwidth]{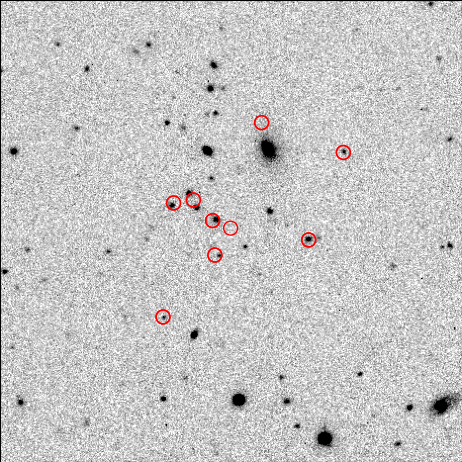} &
\includegraphics[width=0.31\textwidth]{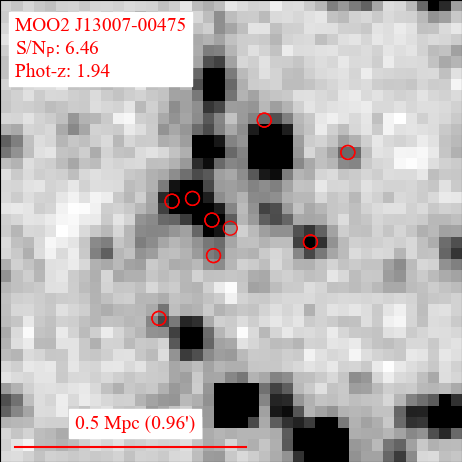}
\end{tabularx}
\caption{\label{fig:pic1}LS DR9 \textit{grz} (left), $K_s$ (middle), and \textit{WISE} W1 (right) images of four clusters candidates at $z$ $>1.5$ for which we have follow-up imaging. The size of the images is $1 \text{~Mpc} \times 1 \text{~Mpc}$. The circles are objects with integrated PDF $>0.2$. The $K_s$ images were obtained from Palomar. In each panel, we provide the name of the cluster candidate, S/N$_{\rm P}$, and its photometric redshift.}
\end{center}
\end{figure*}

A campaign to obtain follow-up near-infrared images of MaDCoWS2 cluster candidates at photometric $z>1.5$ is underway with the FourStar infrared camera \citep{2013Persson} on the Magellan Baade $6.5$-m telescope at Las Campanas Observatory and with the Wide Field Infrared Camera \citep[WIRC][]{2003Wilson} on the $200$-inch Hale Telescope at Palomar Observatory. Full results will be presented in future publications, but in Figure \ref{fig:pic1} we show 1 Mpc cutouts of $K_s$ images for four of these candidates from Palomar. WIRC uses a $2048 \times 2048$ HgCdTe detector with $0\farcs25$ pixels providing a $8.7$ arcmin field of view. The $K_s$ images for MOO2 J03233-06451 ($z_{\rm phot}$ $=1.56$) and MOO2 J09232+10218 ($z_{\rm phot}$ $=1.72$) were obtained on UT 2024 January 14 in moderate cirrus and $\sim 1\farcs5$ seeing, at airmass $1.5$ and $1.1$ respectively. MOO2 J15144+12233 ($z_{\rm phot}$ $=1.63$) was observed on UT 2024 March 9 under clear conditions with $0\farcs7$ seeing at airmass $1.1$, and MOO2 J13007-00475 ($z_{\rm phot}$ $=1.94$) was observed on UT 2024 March 10 under light cirrus with $0\farcs7$ seeing at airmass $1.3$. 
 
In all cases the images were obtained using multiple repeats of a 16-point dither pattern on a grid with $10$ arcsec spacing rotated slightly from the detector array axes to sample independent detector rows and columns, with coadded images at each dither location providing a minute of exposure time per saved frame. The 16-point pattern starting position was offset by a few arcsec to a new value before repeating it to sample additional detector locations. For MOO2 J03233-06451 we coadded six 10-second exposures at each dither location, obtaining a total of $48$ one minute frames. For MOO2 J13007-00475 we coadded five 12-second exposures at each dither, obtaining a total of exposure time $64$ minutes. For MOO2 J09232+10218 and MOO2 J15144+12233 we coadded four 15-second exposures per dither, obtaining total exposure times of $32$ and $24$ minutes respectively. We combined the frames using the dimsum package in IRAF \citep{1995Stanford} to generate the $K_s$ cutouts shown in the middle panels of Figure \ref{fig:pic1}. The left panels show matching cutouts of the combined DECaLS $g$, $r$, and $z$ images, and the right panels show cutouts of the \textit{WISE} W1 images candidates
for the four MaDCoWS2 clusters. Small circles identify objects with integrated PDF $> 0.2$ based on $grzW1W2$ photometry. We define integrated PDF as the integration of the PDF within $\pm0.1(1+z_{\rm cluster})$.

\section{Survey Characterization} \label{sec:survey characterization}
To examine the fidelity of our full survey, we conduct the same analysis as in DR1. We crossmatch MaDCoWS2 cluster candidates to clusters from external catalogs in the area of our search using the matching method described in \S \ref{subsec:crossmatch cluster}. We investigate the fraction of rediscovered clusters, the cluster property relations, the redshift comparison, and positional offsets. 

Following the procedure described in \S 7 of \cite{2024Thongkham}, we remove any clusters from external catalogs that are near/under MaDCoWS2 masks and edges. Specifically, we require the fraction of area masked $f_{\rm area}$ within $30^{\prime\prime}$ and $120^{\prime\prime}$ of the cluster position to be $<0.4$. This ensures that no mismatch exists due to proximity to masks and edges in MaDCoWS2.

\subsection{Existing Cluster Catalogs} \label{subsec:comparison}
The search area of the full release of MaDCoWS2 overlaps with multiple surveys. We choose to compare with the following six wide-area surveys with which we overlap: four intracluster medium (ICM) based surveys, and two optical/infrared cluster surveys. The ICM-based surveys include ACT DR5 \citep{2021Hilton}, SPT \citep{2015Bleem,2019Bocquet,2020Huang,2020Bleem,2024Bleem}, PSZ2 \citep{2016Planck}, and eRASS \citep{2024Bulbul}.
The optical and infrared cluster surveys that we compare to in the area are the redMaPPer catalog \citep{2016Rykoff} and the first MaDCoWS catalog \citep{2019Gonzalez}. This comparison sample is not exhaustive but is intended to be representative. For additional large and relevant cluster catalogs, we refer the reader to the work of \cite{2021Wen,2022Wen,2024Wen,2023Klein,2024KleinA&A,2024Klein}. In contrast to the usage of deep and small catalogs in DR1, DR2 utilizes external catalogs with larger overlapped area for bigger sample size of comparison. The larger area of DR2 increases the number of clusters we compared with for ACT, reMapPPer, and MaDCoWS from $553$, $3071$, $154$ in DR1 to $2232$, $12209$, $665$ in DR2, respectively. This larger sample allows for more robust characterization of DR2. The descriptions of ACT, redMaPPer, and MaDCoWS are available in DR1 and \cite{2021Hilton}, \cite{2016Rykoff}, \cite{2019Gonzalez}, respectively, and are also described in DR1 \citep{2024Thongkham}. We briefly describe the  surveys we did not include in DR1 below.

\subsubsection{SPT} \label{subsec:SPT}
The SPT-SZ 2500d \citep{2015Bleem,2019Bocquet}, SPT-ECS \citep{2020Bleem}, SPTpol 100d \citep{2020Huang}, and SPTpol 500d \citep{2024Bleem} SZ cluster catalogs detect clusters using the thermal SZ signature at $95$ GHz and $150$ GHz. The SPT-SZ 2500d catalog covers $2500$ deg$^2$ in $300^\circ < \text{RA} < 105^\circ$ and $-65^\circ < \text{Dec} < -40^\circ$. Its fiducial depth in the $150$ GHz band is $\leq 18$ $\mu$K arcminute. The SPT-ECS catalog covers $330^\circ < \text{RA} < 90^\circ$ and $-40^\circ < \text{Dec} < -20^\circ$ and $150^\circ < \text{RA} < 210^\circ$ and $-30^\circ < \text{Dec} < -20^\circ$. The noise level in tbe $150$ GHz band is $~30-40$ $\mu$K arcminute. The SPTpol 100d is in $345^\circ < \text{RA} < 0^\circ$ and $-60^\circ < \text{Dec} < -50^\circ$ with a depth level of of $65$ $\mu$K arcminute in the $150$ GHz band. Lastly, SPTpol 500d is centered at RA $=0^\circ$ and Dec$=-57.5^\circ$. The depth of its SZ map is $5.3$ $\mu$K arcminute in the $150$ GHz band. The clusters in these catalogs are confirmed by optical/NIR imaging and have photometric/spectroscopic redshifts. We use the matching criteria defined in \S \ref{subsec:crossmatch cluster} to combine the four catalogs into a catalog consisting of $537$ cluster in the area overlapped with MaDCOWS2. The combined catalog covers $5270$ deg$^2$.

\subsubsection{PSZ2} \label{subsec:psz2}
The PSZ2 survey detects galaxy clusters from the SZ signal using six maps from 100 to 857 GHz \citep{2016Planck}. The catalog consists of $1653$ detections, with $1203$ confirmed by external X-ray, SZ, and infrared data. Redshifts exist for $1094$ of these candidates. The effective area of PSZ2 is $83.6\%$ of the sky. The lower limit on the purity of the catalog is $83\%$ based on comparison with simulated $Planck$ data. 

\subsubsection{eRASS} \label{subsec:eRASS}
The eRASS \citep{2024Merloni} cluster catalog \citep{2024Bulbul,2024Kluge} detects galaxy clusters and groups as extended X-ray sources at $0.2-2.3$ keV using eROSITA onboard the \textit{Russian-German Spectrum-Roentgen-Gamma (SRG)} observatory. The catalog provides $12,247$ optically confirmed clusters at $0.003 <$ $z$ $< 1.32$ over the Western Galactic hemisphere ($359.944 > l >179.944$), which covers $13,116$ deg$^2$. The catalog provides mass $M_{\rm 500c}$ estimated from the eROSITA cosmology pipeline assuming the best-fit relation between the weak-lensing shear measurements and count-rate.

\subsection{Cross-matching Catalogs} \label{subsec:crossmatch cluster}
We use a similar matching method as the one in \cite{2024Thongkham}. Specifically, we find a match for each MaDCoWS2 cluster candidate in order of descending S/N$_{\rm P}$. The matching radius is $1$ Mpc with a redshift window of $\pm \delta z/(1+z)=0.2$. In the case of multiple matches, only the cluster with the largest mass is considered a match. No external cluster is permitted to be matched to multiple MaDCoWS2 cluster candidates.

\subsection{Fraction of Clusters Rediscovered} \label{subsec:rediscovered fraction}

\begin{deluxetable*}{cccccc}[htbp]
\tablecaption{Fit values from the error functions in Figure \ref{fig:completeness} and the total rediscovered fractions. \label{tab:completeness}}
\tablehead{\colhead{Quantity} & \colhead{$\mu$} & \colhead{$\sigma$} & \colhead{$\mu $}& \colhead{$\sigma$} & \colhead{Total $f_{RD}$} \\ 
\colhead{} & \colhead{S/N$_{\rm P}$ $\geq5$} & \colhead{S/N$_{\rm P}$ $\geq5$} & \colhead{S/N$_{\rm P}$ $\geq7$} & \colhead{S/N$_{\rm P}$ $\geq7$} & \colhead{S/N$_{\rm P}$ $\geq5$} } 
\startdata
log$_{10}(M_{\rm 500c}^{\rm ACT})$ & 13.89$_{-0.06}^{+0.05}$ & 0.46$\pm$0.04 & 14.29$_{-0.03}^{+0.02}$ & 0.33$_{-0.03}^{+0.04}$ & 0.91 \\
log$_{10}(M_{\rm 500c}^{\rm SPT})$ & 14.18$_{-0.1}^{+0.06}$ & 0.28$_{-0.07}^{+0.09}$ & 14.39$\pm$0.01 & 0.24$\pm$0.02 & 0.87 \\
log$_{10}(M_{\rm 500c}^{\rm PSZ2})$ & 13.93$_{-2.26}^{+0.34}$ & 0.67$_{-0.3}^{+1.89}$ & 14.18$_{-0.47}^{+0.17}$ & 0.49$_{-0.17}^{+0.43}$ & 0.85 \\
log$_{10}(M_{\rm 500c}^{\rm eRASS})$ & 13.99$_{-0.04}^{+0.03}$ & 0.49$_{-0.04}^{+0.05}$ & 14.29$\pm$0.02 & 0.37$_{-0.02}^{+0.03}$ & 0.72 \\
$\lambda_{\rm redMaPPer}$ & -0.26$\pm$0.54 & 33.51$\pm$1.34 & 36.21$\pm$0.72 & 25.93$\pm$1.65 & 0.86 \\
$\lambda_{\rm MaDCoWS}$ & 25.06$_{-0.87}^{+0.84}$ & 12.77$_{-1.2}^{+1.3}$ & 41.45$_{-1.06}^{+1.16}$ & 10.49$_{-1.14}^{+1.23}$ & 0.59 \\
\enddata
\tablecomments{Best fit $\mu$ and $\sigma$ for the error functions in Figure \ref{fig:completeness}. $\mu$ represents the position on the x-axis with $50\%$ completeness ($f_{RD}=0.5$), and $\sigma$ represents the standard deviation of the error function. The total $f_{RD}$ is the fraction of clusters rediscovered from the total number of clusters. $M_{\rm 500c}^{\text{ACT}}$, $M_{\rm 500c}^{\text{SPT}}$, $M_{\rm 500c}^{\rm PSZ2}$, and $M_{\rm 500c}^{\text{eRASS}}$ are in units of $M_\odot$.}
\end{deluxetable*}

\begin{figure*}[htbp]
\centering
\includegraphics[width=0.9\textwidth]{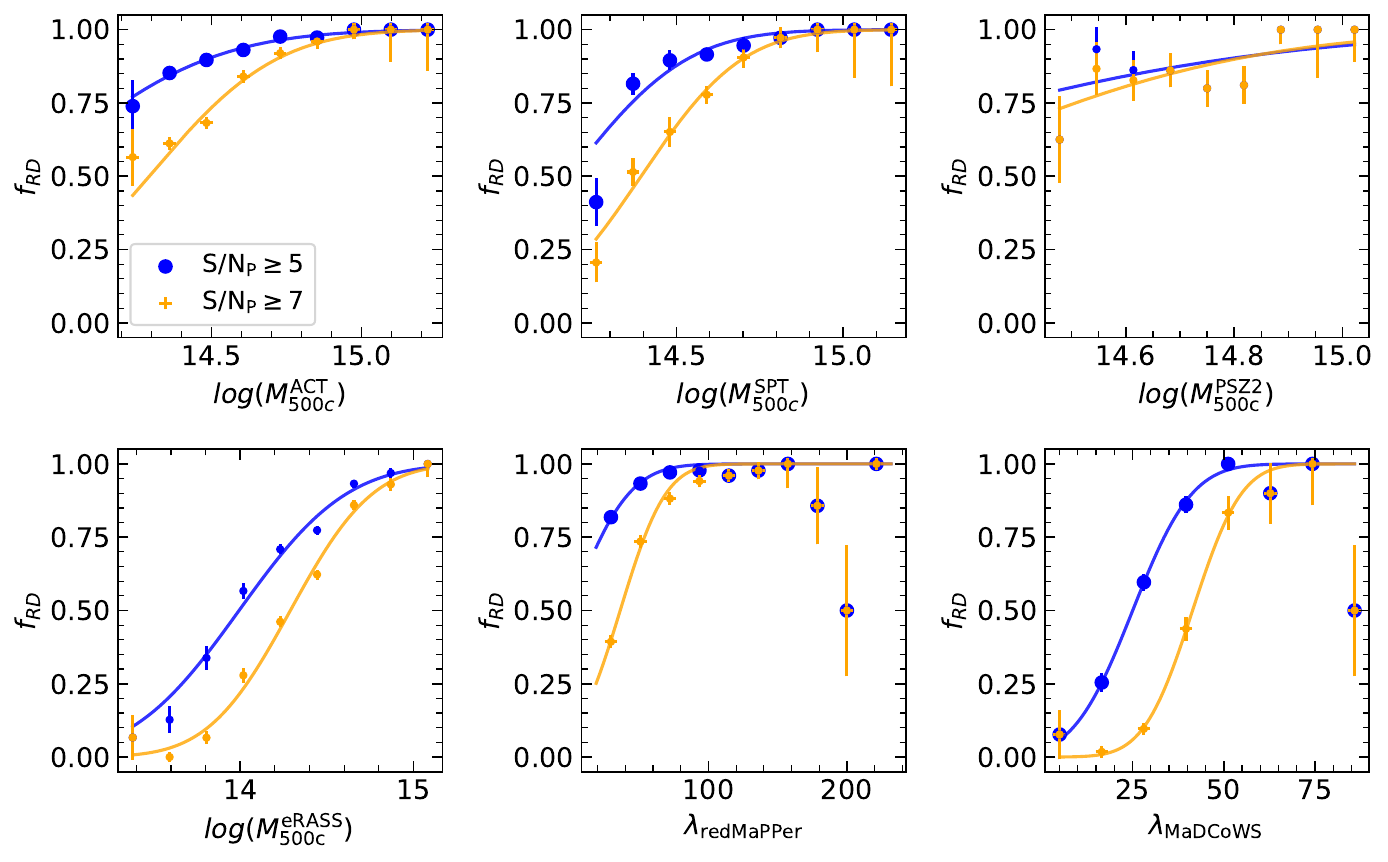}
\caption{$f_{RD}$ as a function of cluster properties for the entire redshift ranges of the external catalogs. We omit the upper error bar when $f_{RD}=1$ for clarity. Blue and yellow dots/lines indicate S/N$_{\rm P}$$\leq5$ and S/N$_{\rm P}$$\leq7$ cutoffs for MaDCoWS2. The solid lines show the function in Equation \ref{eq:errfunc} fitted to the data points. $M_{\rm 500c}^{\rm ACT}$, $M_{\rm 500c}^{\rm SPT}$, $M_{\rm 500c}^{\rm PSZ2}$, $M_{\rm 500c}^{\rm eRASS}$ are in units of $M_\odot$, respectively.}
\label{fig:completeness}
\end{figure*}

\begin{figure*}[htbp]
\centering
\includegraphics[width=0.9\textwidth]{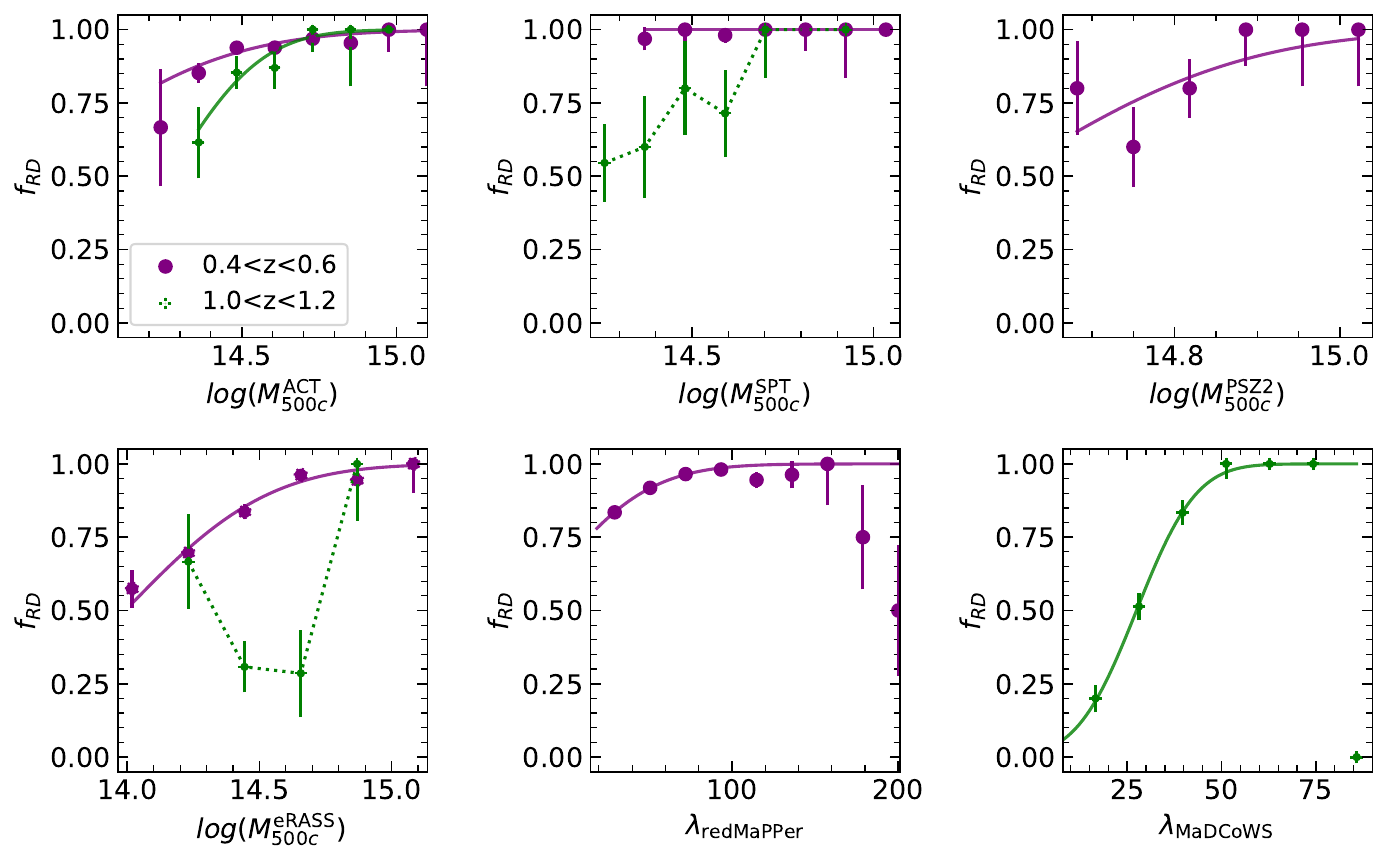}
\caption{The same as Figure \ref{fig:completeness}, but here, the different colors show different redshift ranges for the external catalogs. Purple indicates $0.4<$ $z_{\rm ref}<0.6$, while green indicates $1<$ $z_{\rm ref}<1.2$. This figure adopts an S/N$_{\rm P}$ threshold of S/N$_{\rm P}$ $\geq5$. $M_{\rm 500c}^{\rm ACT}$, $M_{\rm 500c}^{\rm SPT}$, $M_{\rm 500c}^{\rm PSZ2}$, $M_{\rm 500c}^{\rm eRASS}$ are in units of $M_\odot$, respectively.}
\label{fig:completeness2}
\end{figure*}

We assess the fraction of cluster from external catalogs that we rediscover from external catalogs ($f_{RD}$) defined as the number of matches from external catalogs in a given bin divided by the total number of external clusters in that bin. The bins are defined by the mass proxies from the external catalogs. The mass for ACT, SPT, and eRASS is the cluster mass ($M_{\rm 500c}$), estimated from scaling relations, while the mass proxy for redMaPPer and MaDCoWS is the richness ($\lambda$). For ACT, we use the $M_{\rm 500c}$ that is rescaled using the richness-based weak-lensing mass calibration factor of $0.69\pm0.07$ instead of $M_{\rm 500c}$ based on the SZ-mass scaling relation by \cite{2010Arnaud} as done in \cite{2024Thongkham}. Figure \ref{fig:completeness} and \ref{fig:completeness2} show $f_{RD}$ for different S/N$_{\rm P}$ and redshifts. We fit $f_{RD}$ as a function of the mass proxy using a function defined as
\begin{equation}
\label{eq:errfunc}
\begin{split}
    f(x) &= \frac{1}{2}\left(1+\erf{x}\right)\\
    erf(x) &= \frac{2}{\sqrt{\pi}}\int_0^{x}e^{-t^2}dt. 
\end{split}
\end{equation}
where $x=(Q-\mu)/\sqrt{2}\sigma$ and $Q$ denotes the mass estimate or proxy used for binning. The fit $\mu$ and $\sigma$ are shown in Table \ref{tab:completeness}. We note that Equation \ref{eq:errfunc} differs from Equation $5$ from DR1. This function is a cumulative function of a Gaussian distribution derived by integration of the Gaussian distribution \citep{aczel2006complete}. We include $\sqrt{2}$ in DR2 to avoid confusion and more accurately defined the function. We do least squares fitting using the \texttt{lmfit} package \citep{2023Newville}.

At S/N$_{\rm P}$$\geq5$, MaDCoWS2 rediscovers $91\%$, $87\%$, $85\%$, $72\%$, $86\%$ and $59\%$ of cluster candidates from ACT, SPT, PSZ2, eRASS, reMaPPer and MaDCoWS, respectively. In Figure \ref{fig:completeness}, $f_{RD}$ reaches $85\%$ at $\log{M_{\rm 500c}}=$ $14.37$, $14.47$, $14.62$, $14.5$ and $\lambda=$ $34.5$, $38.3$ for ACT, SPT, eRASS, reMaPPer and MaDCoWS, respectively. At $0.4<$ $z_{\rm ref}<0.6$, we detect $93\%$, $99\%$, $79\%$, $80\%$, and $89\%$ of clusters from ACT, SPT, PSZ2, eRASS, and redMaPPer, respectively (Figure \ref{fig:completeness2}). At $1<$ $z_{\rm ref}<1.2$, we find $84\%$, $69\%$, $39\%$, and $55\%$ of clusters from ACT, SPT, eRASS, and MaDCoWS, respectively (Figure \ref{fig:completeness2}). 

As discussed in DR1, while high $f_{RD}$ might be expected for MaDCoWS, the considerable difference between the search method of DR2 and DR1 leads to high scatter in S/N from both surveys. A portion of MaDCoWS clusters may not be detected in MaDCoWS2 due to such scatter.

At $z>1.5$, there are very few spectroscopically confirmed clusters with which we can compare our sample.
In \cite{2024Thongkham} we had three $z>1.5$ systems with which to compare. We rediscovered JKCS 041 
\citep[$z=1.803$,  $M_{200}=4.7\pm{1.5}\times10^{14}\;M_\odot$]{2009Andreon,2023Andreon}, and ClG J0218.3-0510 \citep[$z=1.62$, $M_{200}=0.77\pm{0.38}\times10^{14}\;M_\odot$] {2010Papovich,2012Pierre}. 
The third, SpARCS J022426-032330 \citep[$z=1.633$, $M_{200}\approx2\times10^{14}\;M_\odot$]{2013Muzzin}, was only detected at S/N$_{\rm P}$$=4.5$, so it was not included in our DR1 catalog.

In the expanded area of this release, there are three additional systems of interest. We find a match for  MOO $1723+2946$  \citep[$z=1.603$]{2019Gonzalez} with S/N$_{\rm P}$$=5.86$ at $z_{\rm phot}=1.33$. For SpARCS J022545-035517 \citep[$z=1.60$, $M_{200}\approx1\times10^{14}\;M_\odot$]{2017Noble} we find an overdensity with S/N$_{\rm P}$$=4.02$ at $z_{\rm phot}=1.53$, which is below our S/N$_{\rm P}$ threshold for MaDCoWS2. We find no  significant overdensity at the position of CL J1449+0856  \citep[$z=1.99$, $M_{200}=0.53\pm{0.09}\times10^{14}\;M_\odot$]{2013Gobat}.

It is worth noting that of the six clusters, the only one for which we see no signal is the highest redshift, lowest mass system. Of the other systems, two of the three with $M_{200}\sim1-2\times10^{14}\;M_\odot$ lie below our S/N$_{\rm P}$ threshold. This indicates that our completeness is likely below 50\% at these masses for $z>1.5$.

\subsection{Cluster Properties} \label{subsec:properties}
\begin{figure*}[htbp]
\centering
\includegraphics[width=0.9\textwidth]{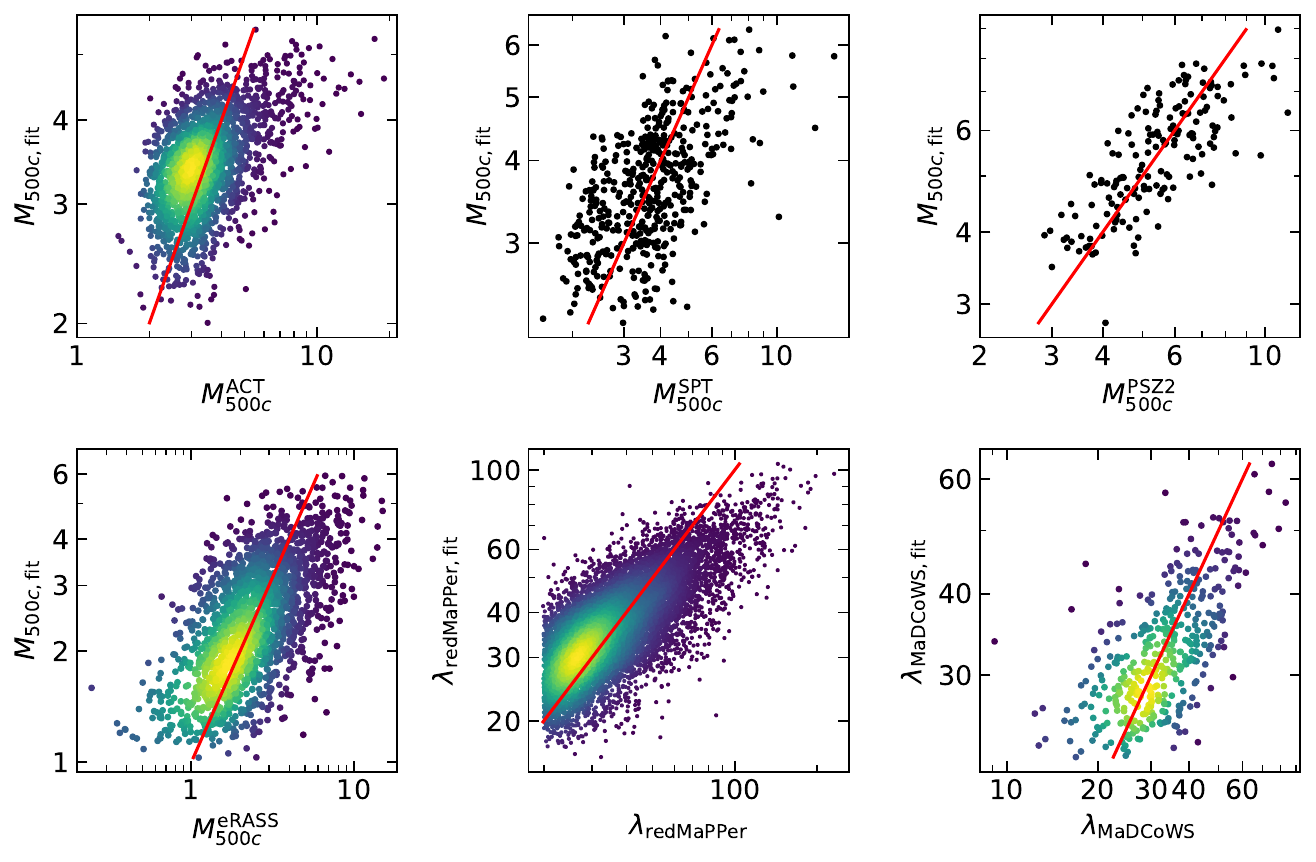}
\caption{Predicted $M_{\rm 500c}$ and $\lambda$ derived from MaDCOWS2 S/N$_{\rm P}$-observable relations versus the published values from external catalogs. These predicted values are derived using Equation \ref{powerlaw} with the best fit values in Table \ref{tab:propfit}. The red lines are the one-to-one lines for each property. $M_{\rm 500c}^{\rm ACT}$, $M_{\rm 500c}^{\rm SPT}$, $M_{\rm 500c}^{\rm PSZ2}$, and $M_{\rm 500c}^{\rm eRASS}$ are given in units of 10$^{14} M_\odot$.}
\label{fig:prop}
\end{figure*}

The MaDCoWS2 S/N$_{\rm P}$ and the cluster properties of external catalogs are compared in Figure \ref{fig:prop}. We fit two power law functions to compare S/N$_{\rm P}$ to these properties using the Markov chain Monte Carlo (MCMC) method from the \texttt{pymc} package \citep{pymc2023}. The power laws follow the equations
\begin{equation}
    \log_{10}\text{S/N$_{\rm P}$} = a + b \log_{10}\text{Q} + c \log_{10}(1+z_{\rm ref})\\
\label{powerlaw}
\end{equation}
\begin{equation}
    \log_{10}\text{Q} = d + e \log_{10}\text{S/N$_{\rm P}$} + f \log_{10}(1+z_{\rm ref})
\label{powerlaw2}
\end{equation}
where $Q$ is any of the quantities from the external catalogs. The fit values are presented in Table \ref{tab:propfit}. The scatter of the S/N$_{\rm P}$ residual is calculated from $\sigma_{\text{\rm S/N res}}$ where S/N$_{\rm P}$ residual$=\text{S/N$_{\rm P}$}-\text{S/N}_\text{fit}$. Similarly, the scatter of the Q residual is determined from $\sigma_{\rm res}$ where residual$=\text{Q}-\text{Q}_\text{fit}$ We obtain fitted cluster properties for MaDCoWS2 by determining $Q$ in equation \ref{powerlaw2}. We display the derived cluster properties S/N$_{\rm P}$ versus the original cluster properties from other catalogs in Figure \ref{fig:prop}.

\begin{deluxetable*}{ccrrrrrrcc}
\tablecaption{Fit values for power law fit to S/N$_{\rm P}$ versus external catalogs quantities. \label{tab:propfit}}
\tablehead{\colhead{Quantity} & \colhead{N(S/N$_{\rm P}$$\geq5$)} & \colhead{$a$} & \colhead{{$b$}} & \colhead{$c$} & \colhead{$d$} & \colhead{{$e$}} & \colhead{$f$} & \colhead{$\sigma_{\rm S/N res}$} & \colhead{$\sigma_{\rm res}$}}
\startdata
$M_{\rm 500c}^{\rm ACT}$ & 2024 & $-4.52\pm$0.23 & 0.39$\pm$0.02 & $-0.90\pm$0.04 & 13.83$\pm$0.03 & 0.62$\pm$0.03 & 0.52$\pm$0.05 & 2.17 & 1.30 \\
$M_{\rm 500c}^{\rm SPT}$ & 468 & $-4.21\pm$0.42 & 0.36$\pm$0.03 & $-0.61\pm$0.08 & 13.91$\pm$0.06 & 0.71$\pm$0.05 & $-0.14\pm$0.09 & 2.15 & 1.30\\
$M_{\rm 500c}^{\rm PSZ2}$ & 163 & $-2.66\pm$0.83 & 0.25$\pm$0.06 & 0.13$\pm$0.20 & 14.01$\pm$0.06 & 0.50$\pm$0.06 & 1.45$\pm$0.14 & 2.86 & 1.11\\
$M_{\rm 500c}^{\rm eRASS}$ & 1848 & $-3.74\pm$0.17 & 0.33$\pm$0.01 & $-0.30\pm$0.05 & 13.19$\pm$0.04 & 1.09$\pm$0.04 & 1.16$\pm$0.08 & 2.22 & 1.49\\
$\lambda_{\rm redMaPPer}$ & 10530 & 0.22$\pm$0.01 & 0.48$\pm$0.01 & $-0.69\pm$0.03 & 0.44$\pm$0.01 & 0.95$\pm$0.01 & 2.14$\pm$0.04 & 1.52 & 13.3\\
$\lambda_{\rm MaDCoWS}$ & 391 & 0.18$\pm$0.08 & 0.58$\pm$0.05 & $-0.83\pm$0.16 & 0.41$\pm$0.08 & 1.02$\pm$0.06 & 0.83$\pm$0.17     & 1.10 & 8.23\\
\enddata
\tablecomments{The fitted values for power law fits between the S/N$_{\rm P}$ of this work and cluster quantities from other surveys (equation \ref{powerlaw} and \ref{powerlaw2} in \S \ref{subsec:properties}). N represents the number of clusters available in each fit. $\sigma_{\text{res}}$ represents the standard deviation of Q residual where Q residual$=\text{Q}-\text{Q}_\text{fit}$while $\sigma_{\text{\rm S/N res}}$ represents the standard deviation of S/N$_{\rm P}$ residual where S/N$_{\rm P}$ residual$=\text{S/N$_{\rm P}$}-\text{S/N}_\text{fit}$. $M_{\rm 500c}^{\rm ACT}$, $M_{\rm 500c}^{\rm SPT}$, $M_{\rm 500c}^{\rm PSZ2}$, and $M_{\rm 500c}^{\rm eRASS}$ are given in units of 10$^{14} M_\odot$.}
\end{deluxetable*}

Similar to the relations in \cite{2024Thongkham}, the galaxy-based catalogs have lower scatter compared to the ICM-based catalogs when fitting mass observables to S/N$_{\rm P}$. We find $\sigma_{\rm res}=1.11,\;1.52,\;2.15,\;2.17,\;2.2$, and $2.86$ for $\lambda_{\text{MaDCoWS}}$, $\lambda_{\rm redMaPPer}$, $M_{\rm 500c}^{\rm SPT}$, $M_{\rm 500c}^{\rm ACT}$, $M_{\rm 500c}^{\rm eRASS}$, and $M_{\rm 500c}^{\rm PSZ2}$, respectively. When using the S/N$_{\rm P}$ to fit for mass, S/N$_{\rm P}$-$M_{\rm 500c}^{\rm PSZ2}$ relation provides the lowest scatter in terms of $M_{\rm 500c}$ compared to other surveys with $M_{\rm 500c}$. The increase in sample size for ACT, redMaPPer, and MaDCoWS results in a better constraint for their parameters, but no major offset is observed.

\subsection{Redshift Comparison} \label{subsec:Redshift comparison}
\begin{figure}[htbp]
\centering
\includegraphics[width=0.9\columnwidth]{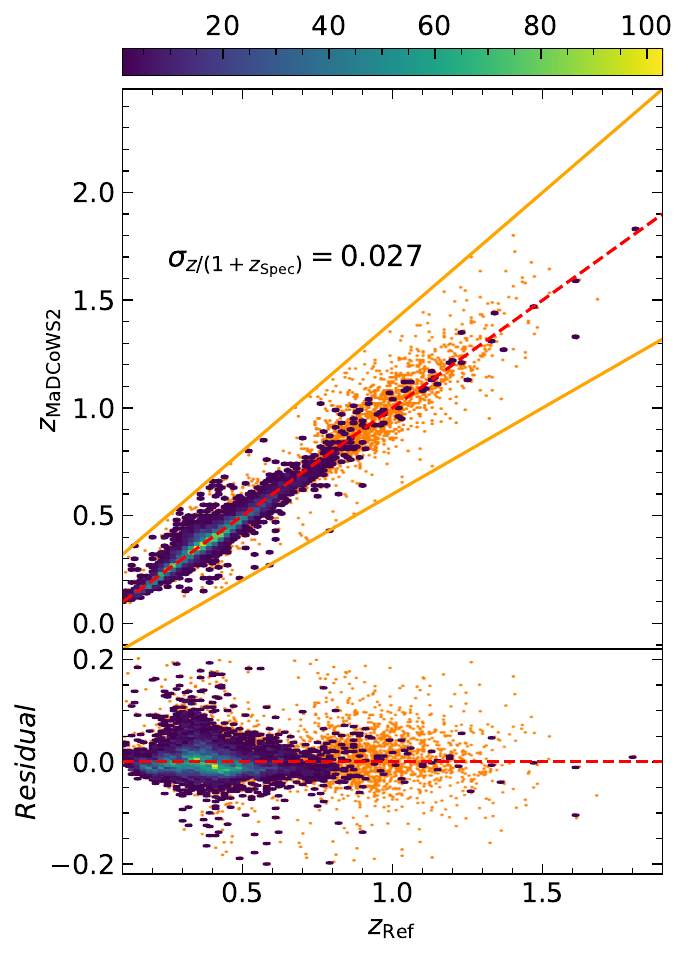}
\caption{Comparison between the photometric redshifts of galaxy cluster candidates in this work and redshifts from external catalogs. Photometric and spectroscopic redshifts are shown as orange points. The spectroscopic redshifts are also overlaid in hexagonal bins with colors indicating the number of clusters in each bin. The hexagonal bin size is $0.02$. The red dashed line indicates a 1:1 relation in the top panel and zero residual in the bottom panel. The bottom panel shows the residual $\delta z /(1+z)$ as a function of redshift.$\sigma_{z/(1+z_{\rm Spec})}$ is the standard deviation of the residual considering only spectroscopic redshifts.}
\label{fig:zcompspec}
\end{figure}

To assess the quality of MaDCOWS2 photometric redshifts, we compare the redshifts from this work to the external redshifts from ACT \citep{2021Hilton}, SPT \citep{2015Bleem,2019Bocquet,2020Bleem,2020Huang,2024Bleem}, PSZ2 \citep{2016Planck}, eRASS \citep{2024Bulbul}, redMaPPer \citep{2016Rykoff}, XXL \citep{2018Adami}, CAMIRA \citep{2014Oguri}, MaDCoWS \citep{2019Gonzalez}, GOGREEN/GCLASS \citep{2017Balogh,2021Balogh}, and clusters from \cite{2009Andreon} and \cite{2010Papovich}. This sample provides $15,626$ total redshifts, including $8111$ spectroscopic redshifts. The compared spectroscopic redshift sample is $2.79$ times larger than the sample of $2904$ in DR1. By the definition of our crossmatching method (see \S \ref{subsec:crossmatch cluster}), the residual $\delta z/(1+z) < 0.2$ for all comparisons. Figure \ref{fig:zcompspec} displays the redshift comparison. Considering all available redshifts, the mean residual of $\overline{\delta z}/(1+z)$ is $0.001$ with the standard deviations of the residual being $\sigma_{z/(1+z)}=$ $0.03$, $0.053$, and $0.031$ at $z<1$, $z>1$, and $z>0.1$, respectively. With only spectroscopic redshifts, the mean residual is $\overline{\delta z}/(1+z_{\rm Spec})=$$0.0002$, while the standard deviations of the residual are $\sigma_{z/(1+z_{\rm Spec})}=$  $0.027$, $0.033$, and $0.027$ at $z<1$, $z>1$, and $z>0.1$, respectively. As in DR1, but now more spectroscopic redshifts, the agreement between our photometric and spectroscopic cluster redshifts is excellent. This will be discussed further in Brodwin et al. (in prep).

\subsection{Offset of the Position of Detections} \label{subsec:offset}
\begin{figure*}[htbp]
\centering
\includegraphics[width=0.9\textwidth]{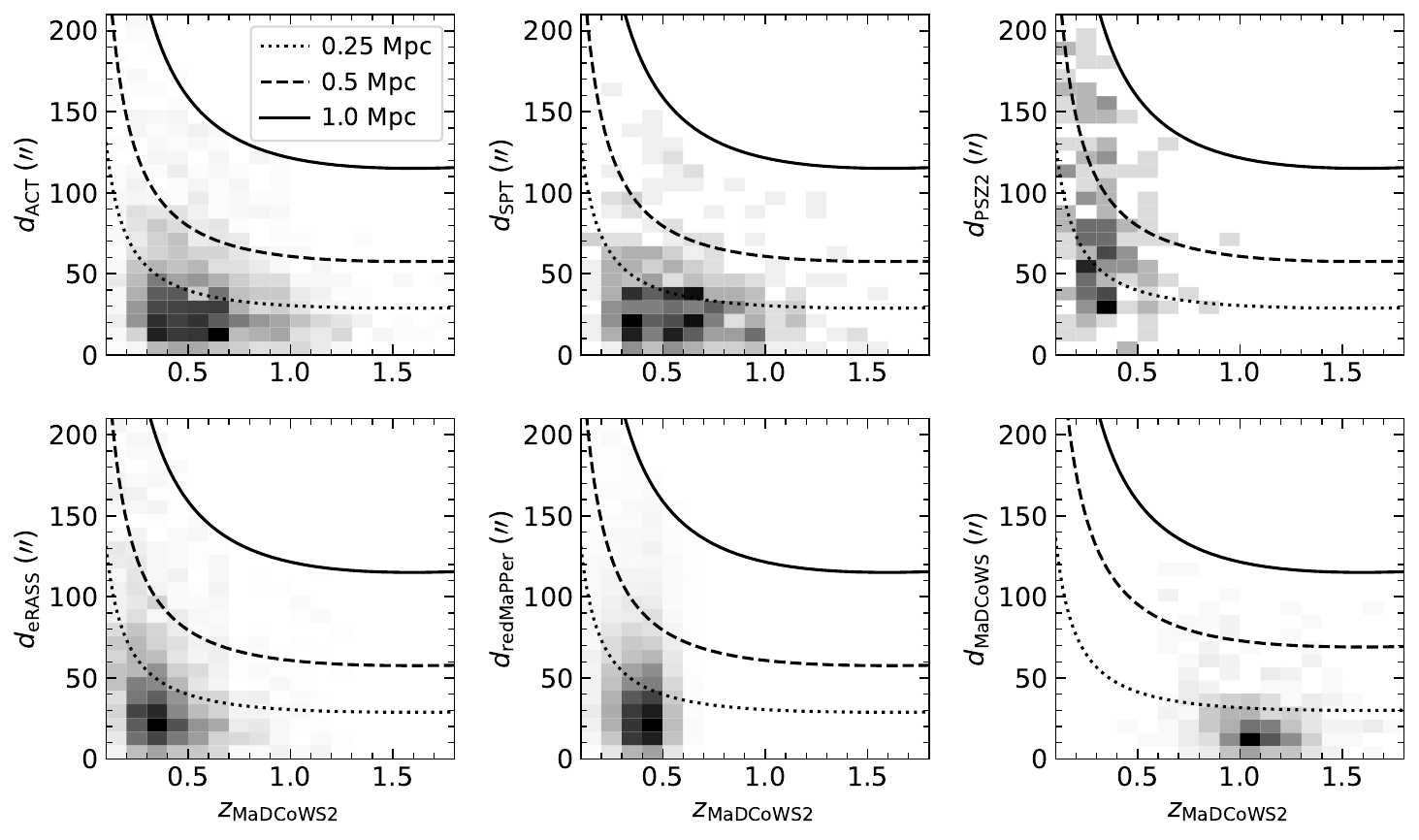}
\includegraphics[width=0.9\textwidth]{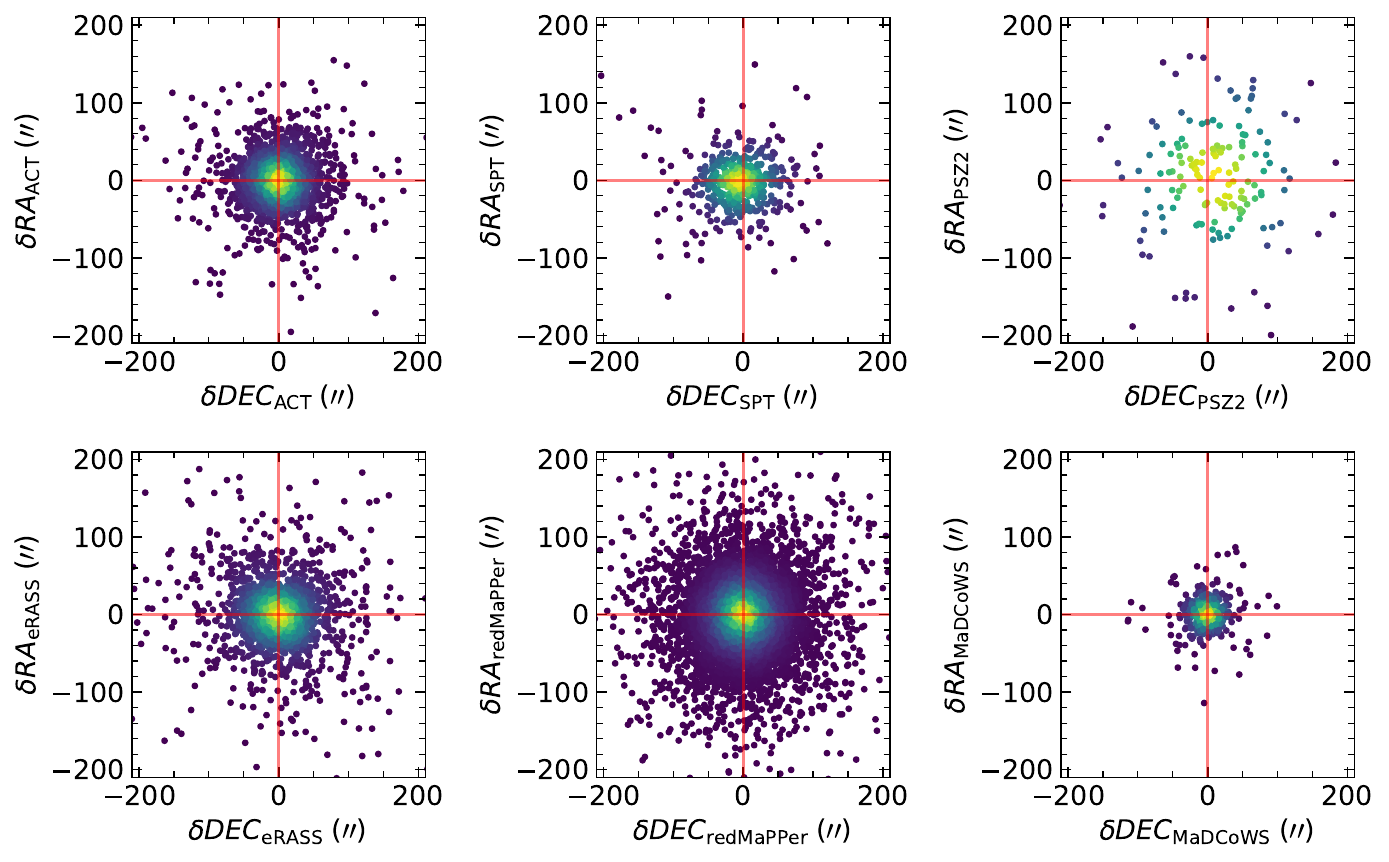}
\caption{Positional offset comparison with external catalogs, where color indicates the density of points. The top two rows show the radial offset $d$ as a function of MaDCoWS2 photometric redshift. The lines denote constant physical radii, as indicated.} 
\label{fig:offset}
\end{figure*}

\begin{figure}[htbp]
\centering
\includegraphics[width=0.9\columnwidth]{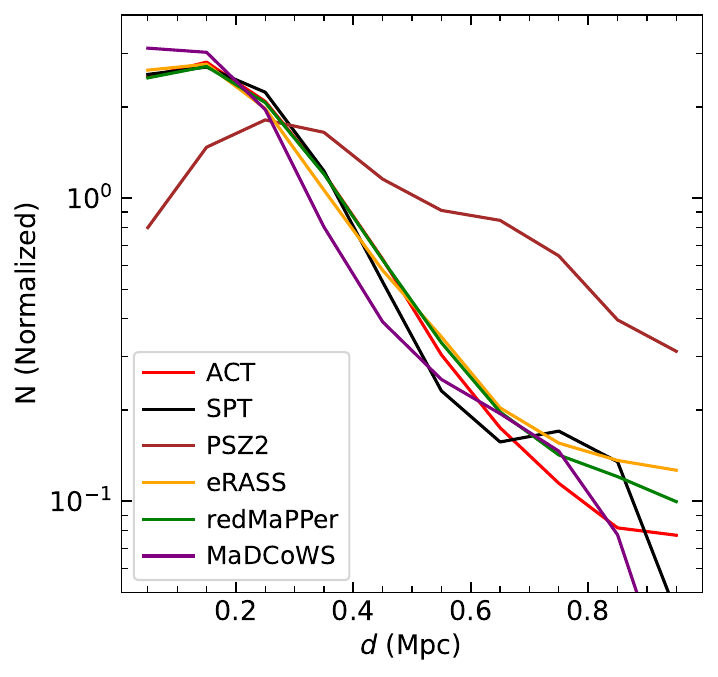}
\caption{Radial offset distribution. The distribution is smoothed by a Gaussian filter with a standard deviation of 0.075 Mpc.}
\label{fig:offset2}
\end{figure}

We display the positional offset between MaDCOWS2 and the external catalogs in Figure \ref{fig:offset} and \ref{fig:offset2}. The median offsets are $0\farcs17,\;6^{\prime\prime},\;14^{\prime\prime},\;0\farcs8,\;0\farcs5$, and $1\farcs3$ for ACT, SPT, PSZ2, eRASS, reMaPPer, and MaDCoWS, respectively. The standard deviations are $33^{\prime\prime},\;43^{\prime\prime},\;102^{\prime\prime},\;58^{\prime\prime},\;44^{\prime\prime}$, and $\;23^{\prime\prime}$, respectively.While the angular offsets vary between surveys due to difference in the redshift distributions of the samples, with the exception of PSZ2, the offset distributions are remarkably similar in physical units. The radial offset distributions shown in Figure \ref{fig:offset2} are well fit by a one-dimensional Gaussian with a standard deviation of 0.2 Mpc. Following the same trend observed in \cite{2024Thongkham}, the peaks of the offset distributions lie below $0.25$ Mpc for all surveys we compared to, with the exception of PSZ2. We also do not observe any systematic offsets between the positions of clusters from MaDCOWS2 and those from external catalogs except PSZ2. This behavior is expected due to the larger uncertainty in the positions of PSZ2 clusters compared to the other catalogs.

\section{PZWav on Simulated Data} \label{sec:simulation}
While the comparison with existing catalogs gives us a sense of the depth of the MaDCoWS2 survey, comparison with simulations enables a measurement of the purity and the completeness as a function of halo mass and redshift.
To assess the performance of our search method on a sample where the true cluster population is known, we therefore ran PZWav on a mock galaxy catalog and compared the resulting cluster catalog with the halo catalog from the simulations.

\subsection{Mock catalog} \label{sec:mock}
The mock galaxy catalog used as input to PZWav is derived from a simulated galaxy sample within a $2$ deg$^2$ light cone from \cite{2023Yung}, who generated this light cone as a forecasting tool for the \textit{Nancy Grace Roman Space Telescope}. The dark matter halos of the light cone are derived from the Small MultiDarkPlanck (SMDPL) simulation from the MultiDark simulation suite \citep{2016Klypin} using the LIGHTCONE package from UNIVERSEMACHINE \citep{2019Behroozi,2020Behroozi}. Merger trees were created from the halos as Monte Carlo realizations with an extended Press-Schechter (EPS)-based method \citep{1993Lacey,1999Somerville,2008Somerville}. The merger history has been proven to qualitatively agree with the N-body simulation though it does not consider environmental effects. The halo mass is defined as the virial mass, the definition adopted from \cite{1998Bryan}. Specifically, the overdensity constant $\Delta_c$ is $18\pi^2$ for a critical universe but has a dependence on cosmology. Santa Cruz semi-analytic models \citep[SAMs;][]{1999Somerville,2015Somerville} forward model the observable properties of the galaxies using detailed prescriptions. The prescriptions are physical processes that are either described analytically or obtained from simulations and observations. We refer the reader to \cite{2023Yung} for the full description of the light cone. A limitation of this light cone is that it contains few massive and few low-redshift ($z$ $\leq0.25$) clusters because of its small solid angle, but the statistics are sufficient near the survey threshold and at higher redshift.

From the full mock galaxy sample, we select only galaxies with W1$<21.4$ (AB) to match the depth of the CatWISE2020 catalog, and require that a galaxy is bright enough to have been detected by DECaLS ($z$ magitude $<23.9$ AB). Because the light cone outputs do not include \textit{WISE} photometry,
we transform IRAC Ch1 and Ch2 from the mock galaxy catalog into \textit{WISE} W1 and W2 using a spectral energy distribution (SED) model given by EzGal \citep{2012Mancone}. To generate the EzGal model, we use a \cite{2003Bruzual} SPS with a Salpeter initial mass function \citep[IMF;][]{1955Salpeter} and exponential star formation history with $\tau=1$ Gyr. After generating the mock data set to be used as input to PZWav, we degraded the data to more closely emulate the real observations. 
In the W1 band, \textit{WISE} has a minor axis Full Width at Half Maximum (FWHM) of 5.6$^{\prime\prime}$ \citep{2012Cutri}. When multiple galaxies lie within one FWHM of one another, we consider these to be blended, combine their flux, and retain only a single merged detection at the position of the brightest galaxy for the mock catalog. 

The PDF for each galaxy is assigned from the ensemble of PDFs associated with galaxies used as inputs in the true MaDCoWS2 catalog. For each galaxy, the PDF is selected based on the MaDCoWS2 photometry. We use a k-nearest neighbor algorithm (k$=1$) to pair mock galaxies to their MaDCoWS2 observational counterparts given their W1-W2 color and redshift. We do not repeat any pairing between a mock galaxy and a MaDCoWS2 galaxy. We input these mock galaxies into PZWav with the same settings as were used in the MaDCoWS2 catalog to create a mock cluster catalog. 

\subsection{Completeness and Purity} \label{sec:purity and completeness}
\begin{figure}[htbp]
\centering
\includegraphics[width=0.9\columnwidth]{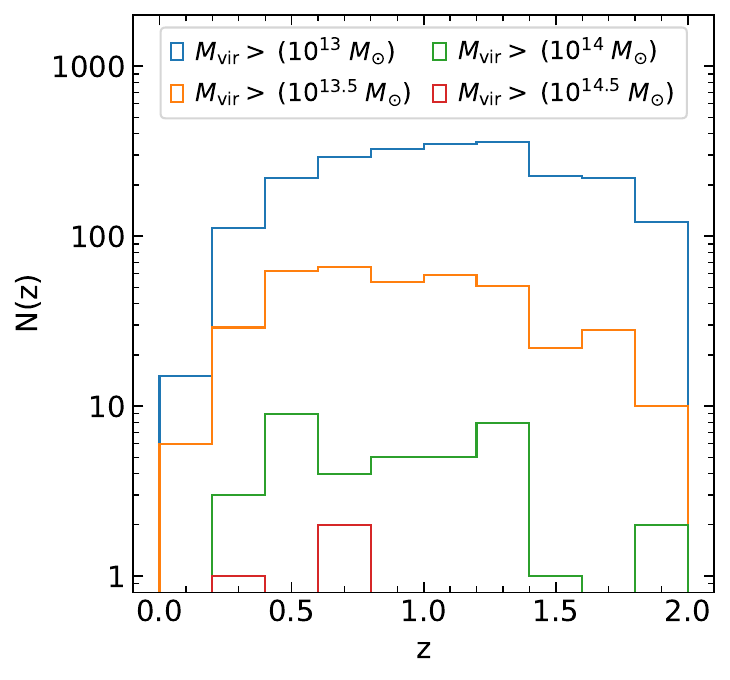}
\caption{Number of simulated halos as a function of redshift for different virial mass thresholds.} 
\label{fig:zdistmock}
\end{figure}

We attempt to estimate the purity and completeness based on cross-matching of the PZWav mock cluster catalog with the simulated halo catalog from \cite{2023Yung}. We define the purity of the catalog as the fraction of detections with a given S/N$_{\rm P}$ for which there is a corresponding halo with $M_{\rm vir}>10^{13} M_\odot$ in the halo catalog that lies within $\delta z=0.2(1+z)$ and within a radius of $1$ Mpc. We use a relatively low mass threshold for the matches in the halo catalog to ensure that the purity is not artificially suppressed due to the detection of lower mass groups below the halo mass threshold. Similarly, we define completeness as the fraction of halos of a given mass that are detected by PZWav. We show the number of halos as function of redshift for different mass thresholds in Figure \ref{fig:zdistmock}. Due to the limited comoving colume, there are only three halos with log$(M_{\rm vir}/M_\odot)>14.5$, and none of these are at $z > 1$, hampering the robustness of this method for estimating completeness.

Nevertheless, we estimate the 50\% completeness virial mass of our S/N$_{\rm P}$$=5$ catalog is log$(M_{\rm vir}/M_\odot) = 14.3\pm0.1$. For virial masses above log$(M_{\rm vir}/M_\odot) = 14.5$, all three of the halos are recovered. Assuming $M_{\rm 500c}/M_{\rm vir}\approx0.6$ based on the estimate from COLOSSUS \citep{2018Diemer} using the concentration model from \cite{2019Diemer}, the 50\% completeness $M_{\rm 500c}$ from our mock cluster search is log$(M_{\rm 500c}/M_\odot) = 14.1\pm0.1$. The 50$\%$ completeness at S/N$_{\rm P}$$=5$ is consistent within the uncertainties with the $M_{\rm 500c}$ values presented in Table \ref{tab:completeness} that are obtained from the rediscovery fractions of the SZ and X-ray selected surveys in \S \ref{subsec:rediscovered fraction}, with the exception of the ACT sample value (log$(M_{\rm 500c}/M_\odot)= 13.89_{-0.06}^{+0.05}$) which is 2.6 sigma lower. 

We measure the purity in a series of redshift bins ($0.1\leq$ $z<0.6$, $0.6\leq$ $z<1.2$, and $1.2\leq$ $z<2$). We find that the sample purity at S/N$_{\rm P}=5$ exceeds $90\%$ in all redshift bins, consistent with the purity level observed in \cite{2019Euclid}. This purity estimate only accounts for contamination due to projection effects. The true purity will be lower due to spurious detections from fragmented big galaxies and bright star artifacts, especially at high redshift and especially if candidates with low CNN probabilities are included.

For the MaDCoWS2 catalog, we set our threshold at S/N$_{\rm P}$$=5$. We expect that contamination due to projection effects will be minimal at this S/N$_{\rm P}$. The 50\% completeness of the catalog at S/N$_{\rm P}$$=5$  is $\sim 2\times10^{14} M_\odot$. Because of the scatter between S/N$_{\rm P}$ and halo mass and the steep slope of the halo mass function, we expect the typical mass of detections to be approximately $10^{14} M_\odot$.

\section{Summary} \label{sec:summary}
The full data release of MaDCoWS2 delivers a catalog $133,036$ cluster candidateswith S/N$_{\rm P}$ $\geq5$ at $0.1\leq z \leq2$. This catalog covers $6498$ deg$^2$ ($8436$ deg$^2$before masking) and spans the majority of the DECaLS area.

To provide a cleaner catalog than the first data release,which covered a smaller area, we train a CNN to identify spurious cluster detections from large galaxies and bright stars. We reach $95\%$ accuracy in classification based on the validation data set.

We perform a mock galaxy cluster search on a $2$ deg$^2$ light-cone using the same search procedure as used for MaDCoWS2. Comparing to the simulated dark matter halo catalog, our search method with a S/N$_{\rm P}$$=5$ threshold provides 50\% completeness at a virial mass of $2.0\pm{0.4}\times10^{14} M_\odot$.

We also compare MaDCoWS2 to ACT, SPT, PSZ2, eRASS, redMaPPer, and MaDCoWS galaxy cluster catalogs using the same analysis conducted in \cite{2024Thongkham}. The larger sample provides a more robust result that is still consistent with our first data release. We find median coordinate offsets less than $250$ kpc. The mean and standard deviation of redshift residuals $\delta z/(1+z_{\rm Ref})$ are below $0.001$ and $0.031$. The relation between MaDCoWS2 S/N$_{\rm P}$ and the richness of optical/infrared surveys produces lower scatter compared to the S/N$_{\rm P}$-mass relation from the SZ and X-ray surveys.

We perform a mock galaxy cluster search on a $2$ deg$^2$ light-cone using the same search procedure as used for MaDCoWS2. Comparing to the simulated dark matter halo catalog, our search method with a S/N$_{\rm P}$$=5$ threshold has minimal contamination due to projection effects. Although the mock catalog has very few halos at this mass,  we estimate 50\% completeness at a virial mass of $\sim 2\times10^{14} M_\odot$, consistent with the recovery fraction found by comparing to other cluster catalogs.

Improvements can still be made for wide-area optical/infrared surveys like MaDCoWS2. Robustly estimating completeness via mock cluster search will require a  larger light cone that contains more clusters at $M_{\rm vir}>10^{14} M_\odot$. Expanding the survey to an even larger area is desirable to discover the most massive clusters at high redshift. The MaDCoWS2 search area could be increased by including the photometric data from the Mayall $z$-band Legacy Survey (MzLS) and Beijing-Arizona Sky Survey (BASS). This would increase the survey area to $>10,000$ deg$^2$ and enable comparison to cluster surveys in the Northern hemisphere. Galaxy photometry with higher resolution and deeper depth is also essential in characterizing high redshift cluster surveys. The incoming data from \textit{Euclid} and the Vera C Rubin Observatory will be valuable in this regard. MaDCoWS2 provides a foundation for such future work by enabling studies of systematics and the properties of wide-area galaxy-based cluster detections.

\acknowledgments
We thank Edward Schlafly of the Space Telescope Science Institute for insights into the DESI Legacy Survey data. We thank Shambhavi Srivastava of the University of Arizona for her insights into the CNN used in this paper.

This material is based upon work supported by the National Science Foundation under Grant No. 2008367. The work of P.E. and D.S. was carried out at the Jet Propulsion Laboratory, California Institute of Technology, under a contract with the National Aeronautics and Space Administration (80NM0018D0004).

This publication makes use of data products from the \textit{Wide-field Infrared Survey Explorer}, which is a joint project of the University of California, Los Angeles, and the Jet Propulsion Laboratory/California Institute of Technology, funded by the National Aeronautics and Space Administration.

This research has made use of the NASA/IPAC Infrared Science Archive, which is funded by the National Aeronautics and Space Administration and operated by the California Institute of Technology.

This research is based upon the Dark Energy Camera Legacy Survey (DECaLS; PIs: David Schlegel and Arjun Dey). The Legacy Surveys imaging of the DESI footprint is supported by the Director, Office of Science, Office of High Energy Physics of the U.S. Department of Energy under Contract No. DE-AC02-05CH1123, by the National Energy Research Scientific Computing Center, a DOE Office of Science User Facility under the same contract; and by the U.S. National Science Foundation, Division of Astronomical Sciences under Contract No.AST-0950952 to NOAO.

This project used data obtained with the Dark Energy Camera (DECam), which was constructed by the Dark Energy Survey (DES) collaboration. Funding for the DES Projects has been provided by the U.S. Department of Energy, the U.S. National Science Foundation, the Ministry of Science and Education of Spain, the Science and Technology Facilities Council of the United Kingdom, the Higher Education Funding Council for England, the National Center for Supercomputing Applications at the University of Illinois at Urbana-Champaign, the Kavli Institute of Cosmological Physics at the University of Chicago, Center for Cosmology and Astro-Particle Physics at the Ohio State University, the Mitchell Institute for Fundamental Physics and Astronomy at Texas A$\&$M University, Financiadora de Estudos e Projetos, Fundação Carlos Chagas Filho de Amparo, Financiadora de Estudos e Projetos, Fundação Carlos Chagas Filho de Amparo à Pesquisa do Estado do Rio de Janeiro, Conselho Nacional de Desenvolvimento Científico e Tecnológico and the Ministério da Ciência, Tecnologia e Inovação, the Deutsche Forschungsgemeinschaft and the Collaborating Institutions in the Dark Energy Survey. The Collaborating Institutions are Argonne National Laboratory, the University of California at Santa Cruz, the University of Cambridge, Centro de Investigaciones Enérgeticas, Medioambientales y Tecnológicas–Madrid, the University of Chicago, University College London, the DES-Brazil Consortium, the University of Edinburgh, the Eidgenössische Technische Hochschule (ETH) Zürich, Fermi National Accelerator Laboratory, the University of Illinois at Urbana-Champaign, the Institut de Ciències de l'Espai (IEEC/CSIC), the Institut de Física d'Altes Energies, Lawrence Berkeley National Laboratory, the Ludwig-Maximilians Universität München and the associated Excellence Cluster Universe, the University of Michigan, the National Optical Astronomy Observatory, the University of Nottingham, the Ohio State University, the University of Pennsylvania, the University of Portsmouth, SLAC National Accelerator Laboratory, Stanford University, the University of Sussex, and Texas A$\&$M University.

The Photometric Redshifts for the Legacy Surveys (PRLS) catalog used in this paper was produced thanks to funding from the U.S. Department of Energy Office of Science, Office of High Energy Physics via grant DE-SC0007914.

The Siena Galaxy Atlas was made possible by funding support from the U.S. Department of Energy, Office of Science, Office of High Energy Physics under Award Number DE-SC0020086 and from the National Science Foundation under grant AST-1616414.

\vspace{5mm}
\software{PZWav, Astropy \citep{astropy:2013, astropy:2018, astropy:2022}, Matplotlib \citep{Hunter:2007}, NumPy \citep{harris2020array}, pandas \citep{the_pandas_development_team_2023_8092754,mckinney-proc-scipy-2010}, lmfit \citep{newville_matthew_2014_11813}, SciPy \citep{2020Virtanen}, fastai \citep{2020Howard}, COLOSSUS \citep{2018Diemer}}

\appendix
\section{Bright Stars} \label{sec:bright stars}
In addition to $8$ stars given in Table 1 of \cite{2024Thongkham}, $27$ stars in the full search area possess diffraction spikes or halos that extend further than the default mask (see Table \ref{Tab:stars}).We create circular mask for the halos and rectangular masks for the diffraction spikes of these stars.

\clearpage

\begin{deluxetable}{cccc}
\tablecaption{Bright Stars Requiring Additional Masking \label{Tab:stars}}
\tablehead{\colhead{Name} & \colhead{RA} & \colhead{Dec} & \colhead{W1$_{\rm AS}$} \\ 
\colhead{} & \colhead{(Deg)} & \colhead{(Deg)} & \colhead{} } 
\startdata
Pi$^1$ Gruis & 335.6913 & -45.9630 & 1.550  \\
Beta Gruis & 340.6531 & -46.8983 & 1.547  \\
Delta$^2$ Gruis & 337.4812 &  -43.7523  & 2.106  \\ 
RS Cancri & 137.6646 &  30.9562 & 3.564  \\
Epsilon Pegasi & 326.0460 &  9.8748 &  2.148  \\
Beta Pegasi & 345.9454 &  28.0821 &  1.981  \\
Pollux & 116.3272 &  28.0259 &  2.087  \\
R Leonis & 146.8890 &  11.4287 &  2.004  \\
Psi Phoenicis & 28.4110  & -46.3030 &  2.225  \\
R Cancri & 124.1412 &  11.7256 &  2.214  \\
EP Aquarii  & 326.6329 & -2.2127 &  2.021  \\
Gamma Eridani & 59.5075 &  -13.5084 &  2.247  \\
CW Leonis & 146.9886 &  13.2801 &  8.649  \\
Tau$^4$ Serpentis & 234.1184  & 15.1032 &  2.455  \\
Tau$^4$ Eridani & 49.8792  & -21.7581 &  2.396 \\
56 Leonis & 164.0062 &  6.1847 &  2.154 \\
Gamma Leonis  & 154.9944 &  19.8412 &  2.362  \\
EK Boötis & 221.5246 &  15.1312 &  2.187 \\
Psi Phoenicis & 28.4110 &  -46.3030 &  2.225 \\
Hamal & 31.7927 &  23.4613 &  2.208  \\
30 Piscium & 0.4910 &  -6.0138 &  2.242 \\
Arcturus & 213.9121 &  19.1812 &  1.438  \\
R Sculptoris & 21.7416 &  -32.5436 &  1.502  \\
Alpha Tucanae  & 334.6237 &  -60.2599 &  1.433  \\
AM Phoenicis & 22.0915 &  -43.3186 &  2.295  \\
\enddata
\tablecomments{The halos or diffraction spikes of these stars extend beyond our default mask sizes and require additional manual masking. W1$_{\rm AS}$ is W1 magnitudes from the All-Sky catalog.}
\end{deluxetable}

\newpage
\clearpage

\bibliography{MaDCoWS2DR2}{}

\begin{thebibliography}{}
\providecommand\natexlab[1]{#1}
\providecommand\JournalTitle[1]{#1}

\bibitem[{{Abell} {et~al.}(1989){Abell}, {Corwin}, \& {Olowin}}]{1989Abell}
{Abell}, G.~O., {Corwin}, Harold~G., J., \& {Olowin}, R.~P. 1989, \href{http://dx.doi.org/10.1086/191333}{\JournalTitle{\apjs}, 70, 1}

\bibitem[{Aczel \& Sounderpandian(2006)}]{aczel2006complete}
Aczel, A., \& Sounderpandian, J. 2006, Complete Business Statistics, Irwin/McGraw-Hill series in operations and decision sciences (Tata McGraw Hill)

\bibitem[{{Adami} {et~al.}(2018){Adami}, {Giles}, {Koulouridis}, {Pacaud}, {Caretta}, {Pierre}, {Eckert}, {Ramos-Ceja}, {Gastaldello}, {Fotopoulou}, {Guglielmo}, {Lidman}, {Sadibekova}, {Iovino}, {Maughan}, {Chiappetti}, {Alis}, {Altieri}, {Baldry}, {Bottini}, {Birkinshaw}, {Bremer}, {Brown}, {Cucciati}, {Driver}, {Elmer}, {Ettori}, {Evrard}, {Faccioli}, {Granett}, {Grootes}, {Guzzo}, {Hopkins}, {Horellou}, {Lef{\`e}vre}, {Liske}, {Malek}, {Marulli}, {Maurogordato}, {Owers}, {Paltani}, {Poggianti}, {Polletta}, {Plionis}, {Pollo}, {Pompei}, {Ponman}, {Rapetti}, {Ricci}, {Robotham}, {Tuffs}, {Tasca}, {Valtchanov}, {Vergani}, {Wagner}, {Willis}, \& {XXL Consortium}}]{2018Adami}
{Adami}, C., {Giles}, P., {Koulouridis}, E., {et~al.} 2018, \href{http://dx.doi.org/10.1051/0004-6361/201731606}{\JournalTitle{\aap}, 620, A5}

\bibitem[{{Aihara} {et~al.}(2018{\natexlab{a}}){Aihara}, {Armstrong}, {Bickerton}, {Bosch}, {Coupon}, {Furusawa}, {Hayashi}, {Ikeda}, {Kamata}, {Karoji}, {Kawanomoto}, {Koike}, {Komiyama}, {Lang}, {Lupton}, {Mineo}, {Miyatake}, {Miyazaki}, {Morokuma}, {Obuchi}, {Oishi}, {Okura}, {Price}, {Takata}, {Tanaka}, {Tanaka}, {Tanaka}, {Uchida}, {Uraguchi}, {Utsumi}, {Wang}, {Yamada}, {Yamanoi}, {Yasuda}, {Arimoto}, {Chiba}, {Finet}, {Fujimori}, {Fujimoto}, {Furusawa}, {Goto}, {Goulding}, {Gunn}, {Harikane}, {Hattori}, {Hayashi}, {He{\l}miniak}, {Higuchi}, {Hikage}, {Ho}, {Hsieh}, {Huang}, {Huang}, {Imanishi}, {Iwata}, {Jaelani}, {Jian}, {Kashikawa}, {Katayama}, {Kojima}, {Konno}, {Koshida}, {Kusakabe}, {Leauthaud}, {Lee}, {Lin}, {Lin}, {Mandelbaum}, {Matsuoka}, {Medezinski}, {Miyama}, {Momose}, {More}, {More}, {Mukae}, {Murata}, {Murayama}, {Nagao}, {Nakata}, {Niida}, {Niikura}, {Nishizawa}, {Oguri}, {Okabe}, {Ono}, {Onodera}, {Onoue}, {Ouchi}, {Pyo}, {Shibuya}, {Shimasaku}, {Simet}, {Speagle}, {Spergel}, {Strauss},
  {Sugahara}, {Sugiyama}, {Suto}, {Suzuki}, {Tait}, {Takada}, {Terai}, {Toba}, {Turner}, {Uchiyama}, {Umetsu}, {Urata}, {Usuda}, {Yeh}, \& {Yuma}}]{2018Aiharab}
{Aihara}, H., {Armstrong}, R., {Bickerton}, S., {et~al.} 2018{\natexlab{a}}, \href{http://dx.doi.org/10.1093/pasj/psx081}{\JournalTitle{\pasj}, 70, S8}

\bibitem[{{Aihara} {et~al.}(2018{\natexlab{b}}){Aihara}, {Arimoto}, {Armstrong}, {Arnouts}, {Bahcall}, {Bickerton}, {Bosch}, {Bundy}, {Capak}, {Chan}, {Chiba}, {Coupon}, {Egami}, {Enoki}, {Finet}, {Fujimori}, {Fujimoto}, {Furusawa}, {Furusawa}, {Goto}, {Goulding}, {Greco}, {Greene}, {Gunn}, {Hamana}, {Harikane}, {Hashimoto}, {Hattori}, {Hayashi}, {Hayashi}, {He{\l}miniak}, {Higuchi}, {Hikage}, {Ho}, {Hsieh}, {Huang}, {Huang}, {Ikeda}, {Imanishi}, {Inoue}, {Iwasawa}, {Iwata}, {Jaelani}, {Jian}, {Kamata}, {Karoji}, {Kashikawa}, {Katayama}, {Kawanomoto}, {Kayo}, {Koda}, {Koike}, {Kojima}, {Komiyama}, {Konno}, {Koshida}, {Koyama}, {Kusakabe}, {Leauthaud}, {Lee}, {Lin}, {Lin}, {Lupton}, {Mandelbaum}, {Matsuoka}, {Medezinski}, {Mineo}, {Miyama}, {Miyatake}, {Miyazaki}, {Momose}, {More}, {More}, {Moritani}, {Moriya}, {Morokuma}, {Mukae}, {Murata}, {Murayama}, {Nagao}, {Nakata}, {Niida}, {Niikura}, {Nishizawa}, {Obuchi}, {Oguri}, {Oishi}, {Okabe}, {Okamoto}, {Okura}, {Ono}, {Onodera}, {Onoue}, {Osato}, {Ouchi},
  {Price}, {Pyo}, {Sako}, {Sawicki}, {Shibuya}, {Shimasaku}, {Shimono}, {Shirasaki}, {Silverman}, {Simet}, {Speagle}, {Spergel}, {Strauss}, {Sugahara}, {Sugiyama}, {Suto}, {Suyu}, {Suzuki}, {Tait}, {Takada}, {Takata}, {Tamura}, {Tanaka}, {Tanaka}, {Tanaka}, {Tanaka}, {Terai}, {Terashima}, {Toba}, {Tominaga}, {Toshikawa}, {Turner}, {Uchida}, {Uchiyama}, {Umetsu}, {Uraguchi}, {Urata}, {Usuda}, {Utsumi}, {Wang}, {Wang}, {Wong}, {Yabe}, {Yamada}, {Yamanoi}, {Yasuda}, {Yeh}, {Yonehara}, \& {Yuma}}]{2018Aihara}
{Aihara}, H., {Arimoto}, N., {Armstrong}, R., {et~al.} 2018{\natexlab{b}}, \href{http://dx.doi.org/10.1093/pasj/psx066}{\JournalTitle{\pasj}, 70, S4}

\bibitem[{{Andreon} {et~al.}(2009){Andreon}, {Maughan}, {Trinchieri}, \& {Kurk}}]{2009Andreon}
{Andreon}, S., {Maughan}, B., {Trinchieri}, G., \& {Kurk}, J. 2009, \href{http://dx.doi.org/10.1051/0004-6361/200912299}{\JournalTitle{\aap}, 507, 147}

\bibitem[{{Andreon} {et~al.}(2023){Andreon}, {Romero}, {Aussel}, {Bhandarkar}, {Devlin}, {Dicker}, {Ladjelate}, {Lowe}, {Mason}, {Mroczkowski}, {Raichoor}, {Sarazin}, \& {Trinchieri}}]{2023Andreon}
{Andreon}, S., {Romero}, C., {Aussel}, H., {et~al.} 2023, \href{http://dx.doi.org/10.1093/mnras/stad1270}{\JournalTitle{\mnras}, 522, 4301}

\bibitem[{{Arnaud} {et~al.}(2010){Arnaud}, {Pratt}, {Piffaretti}, {B{\"o}hringer}, {Croston}, \& {Pointecouteau}}]{2010Arnaud}
{Arnaud}, M., {Pratt}, G.~W., {Piffaretti}, R., {et~al.} 2010, \href{http://dx.doi.org/10.1051/0004-6361/200913416}{\JournalTitle{\aap}, 517, A92}

\bibitem[{{Astropy Collaboration} {et~al.}(2013){Astropy Collaboration}, {Robitaille}, {Tollerud}, {Greenfield}, {Droettboom}, {Bray}, {Aldcroft}, {Davis}, {Ginsburg}, {Price-Whelan}, {Kerzendorf}, {Conley}, {Crighton}, {Barbary}, {Muna}, {Ferguson}, {Grollier}, {Parikh}, {Nair}, {Unther}, {Deil}, {Woillez}, {Conseil}, {Kramer}, {Turner}, {Singer}, {Fox}, {Weaver}, {Zabalza}, {Edwards}, {Azalee Bostroem}, {Burke}, {Casey}, {Crawford}, {Dencheva}, {Ely}, {Jenness}, {Labrie}, {Lim}, {Pierfederici}, {Pontzen}, {Ptak}, {Refsdal}, {Servillat}, \& {Streicher}}]{astropy:2013}
{Astropy Collaboration}, {Robitaille}, T.~P., {Tollerud}, E.~J., {et~al.} 2013, \href{http://dx.doi.org/10.1051/0004-6361/201322068}{\JournalTitle{\aap}, 558, A33}

\bibitem[{{Astropy Collaboration} {et~al.}(2018){Astropy Collaboration}, {Price-Whelan}, {Sip{H{o}}cz}, {G{"u}nther}, {Lim}, {Crawford}, {Conseil}, {Shupe}, {Craig}, {Dencheva}, {Ginsburg}, {Vand erPlas}, {Bradley}, {P{'e}rez-Su{'a}rez}, {de Val-Borro}, {Aldcroft}, {Cruz}, {Robitaille}, {Tollerud}, {Ardelean}, {Babej}, {Bach}, {Bachetti}, {Bakanov}, {Bamford}, {Barentsen}, {Barmby}, {Baumbach}, {Berry}, {Biscani}, {Boquien}, {Bostroem}, {Bouma}, {Brammer}, {Bray}, {Breytenbach}, {Buddelmeijer}, {Burke}, {Calderone}, {Cano Rodr{'i}guez}, {Cara}, {Cardoso}, {Cheedella}, {Copin}, {Corrales}, {Crichton}, {D'Avella}, {Deil}, {Depagne}, {Dietrich}, {Donath}, {Droettboom}, {Earl}, {Erben}, {Fabbro}, {Ferreira}, {Finethy}, {Fox}, {Garrison}, {Gibbons}, {Goldstein}, {Gommers}, {Greco}, {Greenfield}, {Groener}, {Grollier}, {Hagen}, {Hirst}, {Homeier}, {Horton}, {Hosseinzadeh}, {Hu}, {Hunkeler}, {Ivezi{'c}}, {Jain}, {Jenness}, {Kanarek}, {Kendrew}, {Kern}, {Kerzendorf}, {Khvalko}, {King}, {Kirkby}, {Kulkarni}, {Kumar},
  {Lee}, {Lenz}, {Littlefair}, {Ma}, {Macleod}, {Mastropietro}, {McCully}, {Montagnac}, {Morris}, {Mueller}, {Mumford}, {Muna}, {Murphy}, {Nelson}, {Nguyen}, {Ninan}, {N{"o}the}, {Ogaz}, {Oh}, {Parejko}, {Parley}, {Pascual}, {Patil}, {Patil}, {Plunkett}, {Prochaska}, {Rastogi}, {Reddy Janga}, {Sabater}, {Sakurikar}, {Seifert}, {Sherbert}, {Sherwood-Taylor}, {Shih}, {Sick}, {Silbiger}, {Singanamalla}, {Singer}, {Sladen}, {Sooley}, {Sornarajah}, {Streicher}, {Teuben}, {Thomas}, {Tremblay}, {Turner}, {Terr{'o}n}, {van Kerkwijk}, {de la Vega}, {Watkins}, {Weaver}, {Whitmore}, {Woillez}, {Zabalza}, \& {Astropy Contributors}}]{astropy:2018}
{Astropy Collaboration}, {Price-Whelan}, A.~M., {Sip{H{o}}cz}, B.~M., {et~al.} 2018, \href{http://dx.doi.org/10.3847/1538-3881/aabc4f}{\JournalTitle{\aj}, 156, 123}

\bibitem[{{Astropy Collaboration} {et~al.}(2022){Astropy Collaboration}, {Price-Whelan}, {Lim}, {Earl}, {Starkman}, {Bradley}, {Shupe}, {Patil}, {Corrales}, {Brasseur}, {N{"o}the}, {Donath}, {Tollerud}, {Morris}, {Ginsburg}, {Vaher}, {Weaver}, {Tocknell}, {Jamieson}, {van Kerkwijk}, {Robitaille}, {Merry}, {Bachetti}, {G{"u}nther}, {Aldcroft}, {Alvarado-Montes}, {Archibald}, {B{'o}di}, {Bapat}, {Barentsen}, {Baz{'a}n}, {Biswas}, {Boquien}, {Burke}, {Cara}, {Cara}, {Conroy}, {Conseil}, {Craig}, {Cross}, {Cruz}, {D'Eugenio}, {Dencheva}, {Devillepoix}, {Dietrich}, {Eigenbrot}, {Erben}, {Ferreira}, {Foreman-Mackey}, {Fox}, {Freij}, {Garg}, {Geda}, {Glattly}, {Gondhalekar}, {Gordon}, {Grant}, {Greenfield}, {Groener}, {Guest}, {Gurovich}, {Handberg}, {Hart}, {Hatfield-Dodds}, {Homeier}, {Hosseinzadeh}, {Jenness}, {Jones}, {Joseph}, {Kalmbach}, {Karamehmetoglu}, {Ka{l}uszy{'n}ski}, {Kelley}, {Kern}, {Kerzendorf}, {Koch}, {Kulumani}, {Lee}, {Ly}, {Ma}, {MacBride}, {Maljaars}, {Muna}, {Murphy}, {Norman}, {O'Steen},
  {Oman}, {Pacifici}, {Pascual}, {Pascual-Granado}, {Patil}, {Perren}, {Pickering}, {Rastogi}, {Roulston}, {Ryan}, {Rykoff}, {Sabater}, {Sakurikar}, {Salgado}, {Sanghi}, {Saunders}, {Savchenko}, {Schwardt}, {Seifert-Eckert}, {Shih}, {Jain}, {Shukla}, {Sick}, {Simpson}, {Singanamalla}, {Singer}, {Singhal}, {Sinha}, {Sip{H{o}}cz}, {Spitler}, {Stansby}, {Streicher}, {{{S}}umak}, {Swinbank}, {Taranu}, {Tewary}, {Tremblay}, {Val-Borro}, {Van Kooten}, {Vasovi{'c}}, {Verma}, {de Miranda Cardoso}, {Williams}, {Wilson}, {Winkel}, {Wood-Vasey}, {Xue}, {Yoachim}, {Zhang}, {Zonca}, \& {Astropy Project Contributors}}]{astropy:2022}
{Astropy Collaboration}, {Price-Whelan}, A.~M., {Lim}, P.~L., {et~al.} 2022, \href{http://dx.doi.org/10.3847/1538-4357/ac7c74}{\JournalTitle{\apj}, 935, 167}

\bibitem[{{Balogh} {et~al.}(2017){Balogh}, {Gilbank}, {Muzzin}, {Rudnick}, {Cooper}, {Lidman}, {Biviano}, {Demarco}, {McGee}, {Nantais}, {Noble}, {Old}, {Wilson}, {Yee}, {Bellhouse}, {Cerulo}, {Chan}, {Pintos-Castro}, {Simpson}, {van der Burg}, {Zaritsky}, {Ziparo}, {Alonso}, {Bower}, {De Lucia}, {Finoguenov}, {Lambas}, {Muriel}, {Parker}, {Rettura}, {Valotto}, \& {Wetzel}}]{2017Balogh}
{Balogh}, M.~L., {Gilbank}, D.~G., {Muzzin}, A., {et~al.} 2017, \href{http://dx.doi.org/10.1093/mnras/stx1370}{\JournalTitle{\mnras}, 470, 4168}

\bibitem[{{Balogh} {et~al.}(2021){Balogh}, {van der Burg}, {Muzzin}, {Rudnick}, {Wilson}, {Webb}, {Biviano}, {Boak}, {Cerulo}, {Chan}, {Cooper}, {Gilbank}, {Gwyn}, {Lidman}, {Matharu}, {McGee}, {Old}, {Pintos-Castro}, {Reeves}, {Shipley}, {Vulcani}, {Yee}, {Alonso}, {Bellhouse}, {Cooke}, {Davidson}, {De Lucia}, {Demarco}, {Drakos}, {Fillingham}, {Finoguenov}, {Forrest}, {Golledge}, {Jablonka}, {Lambas Garcia}, {McNab}, {Muriel}, {Nantais}, {Noble}, {Parker}, {Petter}, {Poggianti}, {Townsend}, {Valotto}, {Webb}, \& {Zaritsky}}]{2021Balogh}
{Balogh}, M.~L., {van der Burg}, R. F.~J., {Muzzin}, A., {et~al.} 2021, \href{http://dx.doi.org/10.1093/mnras/staa3008}{\JournalTitle{\mnras}, 500, 358}

\bibitem[{{Behroozi} {et~al.}(2019){Behroozi}, {Wechsler}, {Hearin}, \& {Conroy}}]{2019Behroozi}
{Behroozi}, P., {Wechsler}, R.~H., {Hearin}, A.~P., \& {Conroy}, C. 2019, \href{http://dx.doi.org/10.1093/mnras/stz1182}{\JournalTitle{\mnras}, 488, 3143}

\bibitem[{{Behroozi} {et~al.}(2020){Behroozi}, {Conroy}, {Wechsler}, {Hearin}, {Williams}, {Moster}, {Yung}, {Somerville}, {Gottl{\"o}ber}, {Yepes}, \& {Endsley}}]{2020Behroozi}
{Behroozi}, P., {Conroy}, C., {Wechsler}, R.~H., {et~al.} 2020, \href{http://dx.doi.org/10.1093/mnras/staa3164}{\JournalTitle{\mnras}, 499, 5702}

\bibitem[{{Bleem} {et~al.}(2015){Bleem}, {Stalder}, {de Haan}, {Aird}, {Allen}, {Applegate}, {Ashby}, {Bautz}, {Bayliss}, {Benson}, {Bocquet}, {Brodwin}, {Carlstrom}, {Chang}, {Chiu}, {Cho}, {Clocchiatti}, {Crawford}, {Crites}, {Desai}, {Dietrich}, {Dobbs}, {Foley}, {Forman}, {George}, {Gladders}, {Gonzalez}, {Halverson}, {Hennig}, {Hoekstra}, {Holder}, {Holzapfel}, {Hrubes}, {Jones}, {Keisler}, {Knox}, {Lee}, {Leitch}, {Liu}, {Lueker}, {Luong-Van}, {Mantz}, {Marrone}, {McDonald}, {McMahon}, {Meyer}, {Mocanu}, {Mohr}, {Murray}, {Padin}, {Pryke}, {Reichardt}, {Rest}, {Ruel}, {Ruhl}, {Saliwanchik}, {Saro}, {Sayre}, {Schaffer}, {Schrabback}, {Shirokoff}, {Song}, {Spieler}, {Stanford}, {Staniszewski}, {Stark}, {Story}, {Stubbs}, {Vanderlinde}, {Vieira}, {Vikhlinin}, {Williamson}, {Zahn}, \& {Zenteno}}]{2015Bleem}
{Bleem}, L.~E., {Stalder}, B., {de Haan}, T., {et~al.} 2015, \href{http://dx.doi.org/10.1088/0067-0049/216/2/27}{\JournalTitle{\apjs}, 216, 27}

\bibitem[{{Bleem} {et~al.}(2020){Bleem}, {Bocquet}, {Stalder}, {Gladders}, {Ade}, {Allen}, {Anderson}, {Annis}, {Ashby}, {Austermann}, {Avila}, {Avva}, {Bayliss}, {Beall}, {Bechtol}, {Bender}, {Benson}, {Bertin}, {Bianchini}, {Blake}, {Brodwin}, {Brooks}, {Buckley-Geer}, {Burke}, {Carlstrom}, {Rosell}, {Carrasco Kind}, {Carretero}, {Chang}, {Chiang}, {Citron}, {Moran}, {Costanzi}, {Crawford}, {Crites}, {da Costa}, {de Haan}, {De Vicente}, {Desai}, {Diehl}, {Dietrich}, {Dobbs}, {Eifler}, {Everett}, {Flaugher}, {Floyd}, {Frieman}, {Gallicchio}, {Garc{\'\i}a-Bellido}, {George}, {Gerdes}, {Gilbert}, {Gruen}, {Gruendl}, {Gschwend}, {Gupta}, {Gutierrez}, {Halverson}, {Harrington}, {Henning}, {Heymans}, {Holder}, {Hollowood}, {Holzapfel}, {Honscheid}, {Hrubes}, {Huang}, {Hubmayr}, {Irwin}, {James}, {Jeltema}, {Joudaki}, {Khullar}, {Klein}, {Knox}, {Kuropatkin}, {Lee}, {Li}, {Lidman}, {Lowitz}, {MacCrann}, {Mahler}, {Maia}, {Marshall}, {McDonald}, {McMahon}, {Melchior}, {Menanteau}, {Meyer}, {Miquel}, {Mocanu},
  {Mohr}, {Montgomery}, {Nadolski}, {Natoli}, {Nibarger}, {Noble}, {Novosad}, {Padin}, {Palmese}, {Parkinson}, {Patil}, {Paz-Chinch{\'o}n}, {Plazas}, {Pryke}, {Ramachandra}, {Reichardt}, {Remolina Gonz{\'a}lez}, {Romer}, {Roodman}, {Ruhl}, {Rykoff}, {Saliwanchik}, {Sanchez}, {Saro}, {Sayre}, {Schaffer}, {Schrabback}, {Serrano}, {Sharon}, {Sievers}, {Smecher}, {Smith}, {Soares-Santos}, {Stark}, {Story}, {Suchyta}, {Tarle}, {Tucker}, {Vanderlinde}, {Veach}, {Vieira}, {Wang}, {Weller}, {Whitehorn}, {Wu}, {Yefremenko}, \& {Zhang}}]{2020Bleem}
{Bleem}, L.~E., {Bocquet}, S., {Stalder}, B., {et~al.} 2020, \href{http://dx.doi.org/10.3847/1538-4365/ab6993}{\JournalTitle{\apjs}, 247, 25}

\bibitem[{{Bleem} {et~al.}(2024){Bleem}, {Klein}, {Abbot}, {Ade}, {Aguena}, {Alves}, {Anderson}, {Andrade-Oliveira}, {Ansarinejad}, {Archipley}, {Ashby}, {Austermann}, {Bacon}, {Beall}, {Bender}, {Benson}, {Bianchini}, {Bocquet}, {Brooks}, {Burke}, {Calzadilla}, {Carlstrom}, {Carnero Rosell}, {Carretero}, {Chang}, {Chaubal}, {Chiang}, {Chou}, {Citron}, {Corbett Moran}, {Costanzi}, {Constanzi}, {Crawford}, {Crites}, {da Costa}, {de Haan}, {De Vicente}, {Desai}, {Dobbs}, {Doel}, {Everett}, {Ferrero}, {Flaugher}, {Floyd}, {Friedel}, {Frieman}, {Gallicchio}, {Garc'ia-Bellido}, {Gatti}, {George}, {Giannini}, {Grandis}, {Gruen}, {Gruendl}, {Gupta}, {Gutierrez}, {Halverson}, {Hinton}, {Hinton}, {Holder}, {Hollowood}, {Holzapfel}, {Honscheid}, {Hrubes}, {Huang}, {Hubmayr}, {Irwin}, {Mena-Fern{\'a}ndez}, {James}, {K{\'e}ruzor{\'e}}, {Knox}, {Kuehn}, {Lahav}, {Lee}, {Lee}, {Li}, {Lowitz}, {Marshal}, {McDonald}, {McMahon}, {Menanteau}, {Meyer}, {Miquel}, {Mohr}, {Montgomery}, {Myles}, {Natoli}, {Nibarger}, {Noble},
  {Novosad}, {Ogando}, {Padin}, {Patil}, {Pereira}, {Pieres}, {Plazas Malag'on}, {Pryke}, {Reichardt}, {Rodr'iguez-Monroy}, {Romer}, {Ruhl}, {Saliwanchik}, {Salvati}, {Sanchez}, {Saro}, {Schaffer}, {Schrabback}, {Sevilla-Noarbe}, {Sievers}, {Smecher}, {Smith}, {Somboonpanyakul}, {Stalder}, {Stark}, {Suchyta}, {Swanson}, {Tarle}, {To}, {Tucker}, {Veach}, {Vieira}, {Vincenzi}, {Wang}, {Weller}, {Whitehorn}, {Wiseman}, {Wu}, {Yefremenko}, {Zebrowski}, \& {Zhang}}]{2024Bleem}
{Bleem}, L.~E., {Klein}, M., {Abbot}, T.~M.~C., {et~al.} 2024, \href{http://dx.doi.org/10.21105/astro.2311.07512}{\JournalTitle{The Open Journal of Astrophysics}, 7, 13}

\bibitem[{{Bocquet} {et~al.}(2019){Bocquet}, {Dietrich}, {Schrabback}, {Bleem}, {Klein}, {Allen}, {Applegate}, {Ashby}, {Bautz}, {Bayliss}, {Benson}, {Brodwin}, {Bulbul}, {Canning}, {Capasso}, {Carlstrom}, {Chang}, {Chiu}, {Cho}, {Clocchiatti}, {Crawford}, {Crites}, {de Haan}, {Desai}, {Dobbs}, {Foley}, {Forman}, {Garmire}, {George}, {Gladders}, {Gonzalez}, {Grandis}, {Gupta}, {Halverson}, {Hlavacek-Larrondo}, {Hoekstra}, {Holder}, {Holzapfel}, {Hou}, {Hrubes}, {Huang}, {Jones}, {Khullar}, {Knox}, {Kraft}, {Lee}, {von der Linden}, {Luong-Van}, {Mantz}, {Marrone}, {McDonald}, {McMahon}, {Meyer}, {Mocanu}, {Mohr}, {Morris}, {Padin}, {Patil}, {Pryke}, {Rapetti}, {Reichardt}, {Rest}, {Ruhl}, {Saliwanchik}, {Saro}, {Sayre}, {Schaffer}, {Shirokoff}, {Stalder}, {Stanford}, {Staniszewski}, {Stark}, {Story}, {Strazzullo}, {Stubbs}, {Vanderlinde}, {Vieira}, {Vikhlinin}, {Williamson}, \& {Zenteno}}]{2019Bocquet}
{Bocquet}, S., {Dietrich}, J.~P., {Schrabback}, T., {et~al.} 2019, \href{http://dx.doi.org/10.3847/1538-4357/ab1f10}{\JournalTitle{\apj}, 878, 55}

\bibitem[{{Bruzual} \& {Charlot}(2003)}]{2003Bruzual}
{Bruzual}, G., \& {Charlot}, S. 2003, \href{http://dx.doi.org/10.1046/j.1365-8711.2003.06897.x}{\JournalTitle{\mnras}, 344, 1000}

\bibitem[{{Bryan} \& {Norman}(1998)}]{1998Bryan}
{Bryan}, G.~L., \& {Norman}, M.~L. 1998, \href{http://dx.doi.org/10.1086/305262}{\JournalTitle{\apj}, 495, 80}

\bibitem[{{Bulbul} {et~al.}(2024){Bulbul}, {Liu}, {Kluge}, {Zhang}, {Sanders}, {Bahar}, {Ghirardini}, {Artis}, {Seppi}, {Garrel}, {Ramos-Ceja}, {Comparat}, {Balzer}, {B{\"o}ckmann}, {Br{\"u}ggen}, {Clerc}, {Dennerl}, {Dolag}, {Freyberg}, {Grandis}, {Gruen}, {Kleinebreil}, {Krippendorf}, {Lamer}, {Merloni}, {Migkas}, {Nandra}, {Pacaud}, {Predehl}, {Reiprich}, {Schrabback}, {Veronica}, {Weller}, \& {Zelmer}}]{2024Bulbul}
{Bulbul}, E., {Liu}, A., {Kluge}, M., {et~al.} 2024, \href{http://dx.doi.org/10.1051/0004-6361/202348264}{\JournalTitle{\aap}, 685, A106}

\bibitem[{{Coleman} {et~al.}(1980){Coleman}, {Wu}, \& {Weedman}}]{1980CWW}
{Coleman}, G.~D., {Wu}, C.~C., \& {Weedman}, D.~W. 1980, \href{http://dx.doi.org/10.1086/190674}{\JournalTitle{\apjs}, 43, 393}

\bibitem[{{Cutri} {et~al.}(2012){Cutri}, {Wright}, {Conrow}, {Bauer}, {Benford}, {Brandenburg}, {Dailey}, {Eisenhardt}, {Evans}, {Fajardo-Acosta}, {Fowler}, {Gelino}, {Grillmair}, {Harbut}, {Hoffman}, {Jarrett}, {Kirkpatrick}, {Leisawitz}, {Liu}, {Mainzer}, {Marsh}, {Masci}, {McCallon}, {Padgett}, {Ressler}, {Royer}, {Skrutskie}, {Stanford}, {Wyatt}, {Tholen}, {Tsai}, {Wachter}, {Wheelock}, {Yan}, {Alles}, {Beck}, {Grav}, {Masiero}, {McCollum}, {McGehee}, {Papin}, \& {Wittman}}]{2012Cutri}
{Cutri}, R.~M., {Wright}, E.~L., {Conrow}, T., {et~al.} 2012, {Explanatory Supplement to the WISE All-Sky Data Release Products}, Explanatory Supplement to the WISE All-Sky Data Release Products

\bibitem[{{Dey} {et~al.}(2019){Dey}, {Schlegel}, {Lang}, {Blum}, {Burleigh}, {Fan}, {Findlay}, {Finkbeiner}, {Herrera}, {Juneau}, {Landriau}, {Levi}, {McGreer}, {Meisner}, {Myers}, {Moustakas}, {Nugent}, {Patej}, {Schlafly}, {Walker}, {Valdes}, {Weaver}, {Y{\`e}che}, {Zou}, {Zhou}, {Abareshi}, {Abbott}, {Abolfathi}, {Aguilera}, {Alam}, {Allen}, {Alvarez}, {Annis}, {Ansarinejad}, {Aubert}, {Beechert}, {Bell}, {BenZvi}, {Beutler}, {Bielby}, {Bolton}, {Brice{\~n}o}, {Buckley-Geer}, {Butler}, {Calamida}, {Carlberg}, {Carter}, {Casas}, {Castander}, {Choi}, {Comparat}, {Cukanovaite}, {Delubac}, {DeVries}, {Dey}, {Dhungana}, {Dickinson}, {Ding}, {Donaldson}, {Duan}, {Duckworth}, {Eftekharzadeh}, {Eisenstein}, {Etourneau}, {Fagrelius}, {Farihi}, {Fitzpatrick}, {Font-Ribera}, {Fulmer}, {G{\"a}nsicke}, {Gaztanaga}, {George}, {Gerdes}, {Gontcho}, {Gorgoni}, {Green}, {Guy}, {Harmer}, {Hernand ez}, {Honscheid}, {Huang}, {James}, {Jannuzi}, {Jiang}, {Joyce}, {Karcher}, {Karkar}, {Kehoe}, {Kneib}, {Kueter-Young}, {Lan},
  {Lauer}, {Le Guillou}, {Le Van Suu}, {Lee}, {Lesser}, {Perreault Levasseur}, {Li}, {Mann}, {Marshall}, {Mart{\'\i}nez-V{\'a}zquez}, {Martini}, {du Mas des Bourboux}, {McManus}, {Meier}, {M{\'e}nard}, {Metcalfe}, {Mu{\~n}oz-Guti{\'e}rrez}, {Najita}, {Napier}, {Narayan}, {Newman}, {Nie}, {Nord}, {Norman}, {Olsen}, {Paat}, {Palanque-Delabrouille}, {Peng}, {Poppett}, {Poremba}, {Prakash}, {Rabinowitz}, {Raichoor}, {Rezaie}, {Robertson}, {Roe}, {Ross}, {Ross}, {Rudnick}, {Safonova}, {Saha}, {S{\'a}nchez}, {Savary}, {Schweiker}, {Scott}, {Seo}, {Shan}, {Silva}, {Slepian}, {Soto}, {Sprayberry}, {Staten}, {Stillman}, {Stupak}, {Summers}, {Sien Tie}, {Tirado}, {Vargas-Maga{\~n}a}, {Vivas}, {Wechsler}, {Williams}, {Yang}, {Yang}, {Yapici}, {Zaritsky}, {Zenteno}, {Zhang}, {Zhang}, {Zhou}, \& {Zhou}}]{2019Dey}
{Dey}, A., {Schlegel}, D.~J., {Lang}, D., {et~al.} 2019, \href{http://dx.doi.org/10.3847/1538-3881/ab089d}{\JournalTitle{\aj}, 157, 168}

\bibitem[{{Diemer}(2018)}]{2018Diemer}
{Diemer}, B. 2018, \href{http://dx.doi.org/10.3847/1538-4365/aaee8c}{\JournalTitle{\apjs}, 239, 35}

\bibitem[{{Diemer} \& {Joyce}(2019)}]{2019Diemer}
{Diemer}, B., \& {Joyce}, M. 2019, \href{http://dx.doi.org/10.3847/1538-4357/aafad6}{\JournalTitle{\apj}, 871, 168}

\bibitem[{{Ebeling} {et~al.}(2001){Ebeling}, {Edge}, \& {Henry}}]{2001Ebeling}
{Ebeling}, H., {Edge}, A.~C., \& {Henry}, J.~P. 2001, \href{http://dx.doi.org/10.1086/320958}{\JournalTitle{\apj}, 553, 668}

\bibitem[{{Eisenhardt} {et~al.}(2008){Eisenhardt}, {Brodwin}, {Gonzalez}, {Stanford}, {Stern}, {Barmby}, {Brown}, {Dawson}, {Dey}, {Doi}, {Galametz}, {Jannuzi}, {Kochanek}, {Meyers}, {Morokuma}, \& {Moustakas}}]{2008Eisenhardt}
{Eisenhardt}, P. R.~M., {Brodwin}, M., {Gonzalez}, A.~H., {et~al.} 2008, \href{http://dx.doi.org/10.1086/590105}{\JournalTitle{\apj}, 684, 905}

\bibitem[{{Eisenhardt} {et~al.}(2020){Eisenhardt}, {Marocco}, {Fowler}, {Meisner}, {Kirkpatrick}, {Garcia}, {Jarrett}, {Koontz}, {Marchese}, {Stanford}, {Caselden}, {Cushing}, {Cutri}, {Faherty}, {Gelino}, {Gonzalez}, {Mainzer}, {Mobasher}, {Schlegel}, {Stern}, {Teplitz}, \& {Wright}}]{2020Eisenhardt}
{Eisenhardt}, P. R.~M., {Marocco}, F., {Fowler}, J.~W., {et~al.} 2020, \href{http://dx.doi.org/10.3847/1538-4365/ab7f2a}{\JournalTitle{\apjs}, 247, 69}

\bibitem[{{Euclid Collaboration} {et~al.}(2019){Euclid Collaboration}, {Adam}, {Vannier}, {Maurogordato}, {Biviano}, {Adami}, {Ascaso}, {Bellagamba}, {Benoist}, {Cappi}, {D{\'\i}az-S{\'a}nchez}, {Durret}, {Farrens}, {Gonzalez}, {Iovino}, {Licitra}, {Maturi}, {Mei}, {Merson}, {Munari}, {Pell{\'o}}, {Ricci}, {Rocci}, {Roncarelli}, {Sarron}, {Amoura}, {Andreon}, {Apostolakos}, {Arnaud}, {Bardelli}, {Bartlett}, {Baugh}, {Borgani}, {Brodwin}, {Castander}, {Castignani}, {Cucciati}, {De Lucia}, {Dubath}, {Fosalba}, {Giocoli}, {Hoekstra}, {Mamon}, {Melin}, {Moscardini}, {Paltani}, {Radovich}, {Sartoris}, {Schultheis}, {Sereno}, {Weller}, {Burigana}, {Carvalho}, {Corcione}, {Kurki-Suonio}, {Lilje}, {Sirri}, {Toledo-Moreo}, \& {Zamorani}}]{2019Euclid}
{Euclid Collaboration}, {Adam}, R., {Vannier}, M., {et~al.} 2019, \href{http://dx.doi.org/10.1051/0004-6361/201935088}{\JournalTitle{\aap}, 627, A23}

\bibitem[{{Gioia} {et~al.}(1990){Gioia}, {Maccacaro}, {Schild}, {Wolter}, {Stocke}, {Morris}, \& {Henry}}]{1990Gioia}
{Gioia}, I.~M., {Maccacaro}, T., {Schild}, R.~E., {et~al.} 1990, \href{http://dx.doi.org/10.1086/191426}{\JournalTitle{\apjs}, 72, 567}

\bibitem[{{Gladders} \& {Yee}(2005)}]{2005Gladders}
{Gladders}, M.~D., \& {Yee}, H.~K.~C. 2005, \href{http://dx.doi.org/10.1086/427327}{\JournalTitle{\apjs}, 157, 1}

\bibitem[{{Gobat} {et~al.}(2013){Gobat}, {Strazzullo}, {Daddi}, {Onodera}, {Carollo}, {Renzini}, {Finoguenov}, {Cimatti}, {Scarlata}, \& {Arimoto}}]{2013Gobat}
{Gobat}, R., {Strazzullo}, V., {Daddi}, E., {et~al.} 2013, \href{http://dx.doi.org/10.1088/0004-637X/776/1/9}{\JournalTitle{\apj}, 776, 9}

\bibitem[{{Gonzalez}(2014)}]{2014Gonzalez}
{Gonzalez}, A. 2014, in Building the Euclid Cluster Survey - Scientific Program, 7

\bibitem[{{Gonzalez} {et~al.}(2019){Gonzalez}, {Gettings}, {Brodwin}, {Eisenhardt}, {Stanford}, {Wylezalek}, {Decker}, {Marrone}, {Moravec}, {O'Donnell}, {Stalder}, {Stern}, {Abdulla}, {Brown}, {Carlstrom}, {Chambers}, {Hayden}, {Lin}, {Magnier}, {Masci}, {Mantz}, {McDonald}, {Mo}, {Perlmutter}, {Wright}, \& {Zeimann}}]{2019Gonzalez}
{Gonzalez}, A.~H., {Gettings}, D.~P., {Brodwin}, M., {et~al.} 2019, \href{http://dx.doi.org/10.3847/1538-4365/aafad2}{\JournalTitle{\apjs}, 240, 33}

\bibitem[{Harris {et~al.}(2020)Harris, Millman, van~der Walt, Gommers, Virtanen, Cournapeau, Wieser, Taylor, Berg, Smith, Kern, Picus, Hoyer, van Kerkwijk, Brett, Haldane, del R{\'{i}}o, Wiebe, Peterson, G{\'{e}}rard-Marchant, Sheppard, Reddy, Weckesser, Abbasi, Gohlke, \& Oliphant}]{harris2020array}
Harris, C.~R., Millman, K.~J., van~der Walt, S.~J., {et~al.} 2020, \href{http://dx.doi.org/10.1038/s41586-020-2649-2}{\JournalTitle{Nature}, 585, 357}

\bibitem[{{He} {et~al.}(2015){He}, {Zhang}, {Ren}, \& {Sun}}]{2015He}
{He}, K., {Zhang}, X., {Ren}, S., \& {Sun}, J. 2015, \href{http://dx.doi.org/10.48550/arXiv.1512.03385}{\JournalTitle{arXiv e-prints}, arXiv:1512.03385}

\bibitem[{{Hilton} {et~al.}(2021){Hilton}, {Sif{\'o}n}, {Naess}, {Madhavacheril}, {Oguri}, {Rozo}, {Rykoff}, {Abbott}, {Adhikari}, {Aguena}, {Aiola}, {Allam}, {Amodeo}, {Amon}, {Annis}, {Ansarinejad}, {Aros-Bunster}, {Austermann}, {Avila}, {Bacon}, {Battaglia}, {Beall}, {Becker}, {Bernstein}, {Bertin}, {Bhandarkar}, {Bhargava}, {Bond}, {Brooks}, {Burke}, {Calabrese}, {Carrasco Kind}, {Carretero}, {Choi}, {Choi}, {Conselice}, {da Costa}, {Costanzi}, {Crichton}, {Crowley}, {D{\"u}nner}, {Denison}, {Devlin}, {Dicker}, {Diehl}, {Dietrich}, {Doel}, {Duff}, {Duivenvoorden}, {Dunkley}, {Everett}, {Ferraro}, {Ferrero}, {Fert{\'e}}, {Flaugher}, {Frieman}, {Gallardo}, {Garc{\'\i}a-Bellido}, {Gaztanaga}, {Gerdes}, {Giles}, {Golec}, {Gralla}, {Grandis}, {Gruen}, {Gruendl}, {Gschwend}, {Gutierrez}, {Han}, {Hartley}, {Hasselfield}, {Hill}, {Hilton}, {Hincks}, {Hinton}, {Ho}, {Honscheid}, {Hoyle}, {Hubmayr}, {Huffenberger}, {Hughes}, {Jaelani}, {Jain}, {James}, {Jeltema}, {Kent}, {Knowles}, {Koopman}, {Kuehn}, {Lahav},
  {Lima}, {Lin}, {Lokken}, {Loubser}, {MacCrann}, {Maia}, {Marriage}, {Martin}, {McMahon}, {Melchior}, {Menanteau}, {Miquel}, {Miyatake}, {Moodley}, {Morgan}, {Mroczkowski}, {Nati}, {Newburgh}, {Niemack}, {Nishizawa}, {Ogando}, {Orlowski-Scherer}, {Page}, {Palmese}, {Partridge}, {Paz-Chinch{\'o}n}, {Phakathi}, {Plazas}, {Robertson}, {Romer}, {Carnero Rosell}, {Salatino}, {Sanchez}, {Schaan}, {Schillaci}, {Sehgal}, {Serrano}, {Shin}, {Simon}, {Smith}, {Soares-Santos}, {Spergel}, {Staggs}, {Storer}, {Suchyta}, {Swanson}, {Tarle}, {Thomas}, {To}, {Trac}, {Ullom}, {Vale}, {Van Lanen}, {Vavagiakis}, {De Vicente}, {Wilkinson}, {Wollack}, {Xu}, \& {Zhang}}]{2021Hilton}
{Hilton}, M., {Sif{\'o}n}, C., {Naess}, S., {et~al.} 2021, \href{http://dx.doi.org/10.3847/1538-4365/abd023}{\JournalTitle{\apjs}, 253, 3}

\bibitem[{{Howard} \& {Gugger}(2020)}]{2020Howard}
{Howard}, J., \& {Gugger}, S. 2020, \href{http://dx.doi.org/10.48550/arXiv.2002.04688}{\JournalTitle{arXiv e-prints}, arXiv:2002.04688}

\bibitem[{{Huang} {et~al.}(2020){Huang}, {Bleem}, {Stalder}, {Ade}, {Allen}, {Anderson}, {Austermann}, {Avva}, {Beall}, {Bender}, {Benson}, {Bianchini}, {Bocquet}, {Brodwin}, {Carlstrom}, {Chang}, {Chiang}, {Citron}, {Moran}, {Crawford}, {Crites}, {Haan}, {Dobbs}, {Everett}, {Floyd}, {Gallicchio}, {George}, {Gilbert}, {Gladders}, {Guns}, {Gupta}, {Halverson}, {Harrington}, {Henning}, {Hilton}, {Holder}, {Holzapfel}, {Hrubes}, {Hubmayr}, {Irwin}, {Khullar}, {Knox}, {Lee}, {Li}, {Lowitz}, {McDonald}, {McMahon}, {Meyer}, {Mocanu}, {Montgomery}, {Nadolski}, {Natoli}, {Nibarger}, {Noble}, {Novosad}, {Padin}, {Patil}, {Pryke}, {Reichardt}, {Ruhl}, {Saliwanchik}, {Saro}, {Sayre}, {Schaffer}, {Sharon}, {Sievers}, {Smecher}, {Stark}, {Story}, {Tucker}, {Vanderlinde}, {Veach}, {Vieira}, {Wang}, {Whitehorn}, {Wu}, \& {Yefremenko}}]{2020Huang}
{Huang}, N., {Bleem}, L.~E., {Stalder}, B., {et~al.} 2020, \href{http://dx.doi.org/10.3847/1538-3881/ab6a96}{\JournalTitle{\aj}, 159, 110}

\bibitem[{Hunter(2007)}]{Hunter:2007}
Hunter, J.~D. 2007, \href{http://dx.doi.org/10.1109/MCSE.2007.55}{\JournalTitle{Computing in Science \& Engineering}, 9, 90}

\bibitem[{{Klein} {et~al.}(2023){Klein}, {Mohr}, {Bocquet}, {SPT}, \& {collabortions}}]{2023Klein}
{Klein}, M., {Mohr}, J.~J., {Bocquet}, S., {SPT}, t., \& {collabortions}, D. 2023, \href{http://dx.doi.org/10.48550/arXiv.2309.09908}{\JournalTitle{arXiv e-prints}, arXiv:2309.09908}

\bibitem[{{Klein} {et~al.}(2024{\natexlab{a}}){Klein}, {Mohr}, \& {Davies}}]{2024KleinA&A}
{Klein}, M., {Mohr}, J.~J., \& {Davies}, C.~T. 2024{\natexlab{a}}, \href{http://dx.doi.org/10.1051/0004-6361/202451203}{\JournalTitle{\aap}, 690, A322}

\bibitem[{{Klein} {et~al.}(2024{\natexlab{b}}){Klein}, {Mohr}, {Bocquet}, {Aguena}, {Allen}, {Alves}, {Ansarinejad}, {Ashby}, {Bacon}, {Bayliss}, {Benson}, {Bleem}, {Brodwin}, {Brooks}, {Bulbul}, {Burke}, {Canning}, {Carlstrom}, {Rosell}, {Carretero}, {Chang}, {Conselice}, {Costanzi}, {Crites}, {da Costa}, {Pereira}, {Davis}, {De Vicente}, {Desai}, {de Haan}, {Dobbs}, {Doel}, {Ferrero}, {Flores}, {Frieman}, {George}, {Giannini}, {Gladders}, {Gonzalez}, {Grandis}, {Gruen}, {Gruendl}, {Gutierrez}, {Halverson}, {Hinton}, {Holder}, {Hollowood}, {Holzapfel}, {Honscheid}, {Hrubes}, {Huang}, {James}, {Khullar}, {Kim}, {Knox}, {Kraft}, {K{\'e}ruzor{\'e}}, {Lee}, {Luong-Van}, {Mahler}, {Mantz}, {Marrone}, {Marshall}, {McDonald}, {McMahon}, {Mena-Fern{\'a}ndez}, {Menanteau}, {Meyer}, {Miquel}, {Myles}, {Padin}, {Pieres}, {Plazas Malag{\'o}n}, {Pryke}, {Reichardt}, {Reil}, {Roberson}, {Romer}, {Romero}, {Ruhl}, {Saliwanchik}, {Salvati}, {Sanchez}, {Saro}, {Schaffer}, {Schrabback}, {Schubnell}, {Sevilla-Noarbe},
  {Sharon}, {Shirokoff}, {Smith}, {Somboonpanyakul}, {Stalder}, {Stanford}, {Stark}, {Strazzullo}, {Suchyta}, {Swanson}, {Tarle}, {To}, {Vanderlinde}, {Vieira}, {von der Linden}, {Weaverdyck}, {Williamson}, {Wiseman}, \& {Young}}]{2024Klein}
{Klein}, M., {Mohr}, J.~J., {Bocquet}, S., {et~al.} 2024{\natexlab{b}}, \href{http://dx.doi.org/10.1093/mnras/stae1359}{\JournalTitle{\mnras}, 531, 3973}

\bibitem[{{Kluge} {et~al.}(2024){Kluge}, {Comparat}, {Liu}, {Balzer}, {Bulbul}, {Ider Chitham}, {Ghirardini}, {Garrel}, {Bahar}, {Artis}, {Bender}, {Clerc}, {Dwelly}, {Fabricius}, {Grandis}, {Hern{\'a}ndez-Lang}, {Hill}, {Joshi}, {Lamer}, {Merloni}, {Nandra}, {Pacaud}, {Predehl}, {Ramos-Ceja}, {Reiprich}, {Salvato}, {Sanders}, {Schrabback}, {Seppi}, {Zelmer}, {Zenteno}, \& {Zhang}}]{2024Kluge}
{Kluge}, M., {Comparat}, J., {Liu}, A., {et~al.} 2024, \href{http://dx.doi.org/10.1051/0004-6361/202349031}{\JournalTitle{\aap}, 688, A210}

\bibitem[{{Klypin} {et~al.}(2016){Klypin}, {Yepes}, {Gottl{\"o}ber}, {Prada}, \& {He{\ss}}}]{2016Klypin}
{Klypin}, A., {Yepes}, G., {Gottl{\"o}ber}, S., {Prada}, F., \& {He{\ss}}, S. 2016, \href{http://dx.doi.org/10.1093/mnras/stw248}{\JournalTitle{\mnras}, 457, 4340}

\bibitem[{{Lacey} \& {Cole}(1993)}]{1993Lacey}
{Lacey}, C., \& {Cole}, S. 1993, \href{http://dx.doi.org/10.1093/mnras/262.3.627}{\JournalTitle{\mnras}, 262, 627}

\bibitem[{{Liu} {et~al.}(2022){Liu}, {Bulbul}, {Ghirardini}, {Liu}, {Klein}, {Clerc}, {{\"O}zsoy}, {Ramos-Ceja}, {Pacaud}, {Comparat}, {Okabe}, {Bahar}, {Biffi}, {Brunner}, {Br{\"u}ggen}, {Buchner}, {Ider Chitham}, {Chiu}, {Dolag}, {Gatuzz}, {Gonzalez}, {Hoang}, {Lamer}, {Merloni}, {Nandra}, {Oguri}, {Ota}, {Predehl}, {Reiprich}, {Salvato}, {Schrabback}, {Sanders}, {Seppi}, \& {Thibaud}}]{2022Liu}
{Liu}, A., {Bulbul}, E., {Ghirardini}, V., {et~al.} 2022, \href{http://dx.doi.org/10.1051/0004-6361/202141120}{\JournalTitle{\aap}, 661, A2}

\bibitem[{{Mancone} \& {Gonzalez}(2012)}]{2012Mancone}
{Mancone}, C.~L., \& {Gonzalez}, A.~H. 2012, \href{http://dx.doi.org/10.1086/666502}{\JournalTitle{\pasp}, 124, 606}

\bibitem[{{Marocco} {et~al.}(2021){Marocco}, {Eisenhardt}, {Fowler}, {Kirkpatrick}, {Meisner}, {Schlafly}, {Stanford}, {Garcia}, {Caselden}, {Cushing}, {Cutri}, {Faherty}, {Gelino}, {Gonzalez}, {Jarrett}, {Koontz}, {Mainzer}, {Marchese}, {Mobasher}, {Schlegel}, {Stern}, {Teplitz}, \& {Wright}}]{2021Marocco}
{Marocco}, F., {Eisenhardt}, P. R.~M., {Fowler}, J.~W., {et~al.} 2021, \href{http://dx.doi.org/10.3847/1538-4365/abd805}{\JournalTitle{\apjs}, 253, 8}

\bibitem[{{Mehrtens} {et~al.}(2012){Mehrtens}, {Romer}, {Hilton}, {Lloyd-Davies}, {Miller}, {Stanford}, {Hosmer}, {Hoyle}, {Collins}, {Liddle}, {Viana}, {Nichol}, {Stott}, {Dubois}, {Kay}, {Sahl{\'e}n}, {Young}, {Short}, {Christodoulou}, {Watson}, {Davidson}, {Harrison}, {Baruah}, {Smith}, {Burke}, {Mayers}, {Deadman}, {Rooney}, {Edmondson}, {West}, {Campbell}, {Edge}, {Mann}, {Sabirli}, {Wake}, {Benoist}, {da Costa}, {Maia}, \& {Ogando}}]{2012Mehrtens}
{Mehrtens}, N., {Romer}, A.~K., {Hilton}, M., {et~al.} 2012, \href{http://dx.doi.org/10.1111/j.1365-2966.2012.20931.x}{\JournalTitle{\mnras}, 423, 1024}

\bibitem[{{Merloni} {et~al.}(2024){Merloni}, {Lamer}, {Liu}, {Ramos-Ceja}, {Brunner}, {Bulbul}, {Dennerl}, {Doroshenko}, {Freyberg}, {Friedrich}, {Gatuzz}, {Georgakakis}, {Haberl}, {Igo}, {Kreykenbohm}, {Liu}, {Maitra}, {Malyali}, {Mayer}, {Nandra}, {Predehl}, {Robrade}, {Salvato}, {Sanders}, {Stewart}, {Tub{\'\i}n-Arenas}, {Weber}, {Wilms}, {Arcodia}, {Artis}, {Aschersleben}, {Avakyan}, {Aydar}, {Bahar}, {Balzer}, {Becker}, {Berger}, {Boller}, {Bornemann}, {Br{\"u}ggen}, {Brusa}, {Buchner}, {Burwitz}, {Camilloni}, {Clerc}, {Comparat}, {Coutinho}, {Czesla}, {Dannhauer}, {Dauner}, {Dauser}, {Dietl}, {Dolag}, {Dwelly}, {Egg}, {Ehl}, {Freund}, {Friedrich}, {Gaida}, {Garrel}, {Ghirardini}, {Gokus}, {Gr{\"u}nwald}, {Grandis}, {Grotova}, {Gruen}, {Gueguen}, {H{\"a}mmerich}, {Hamaus}, {Hasinger}, {Haubner}, {Homan}, {Ider Chitham}, {Joseph}, {Joyce}, {K{\"o}nig}, {Kaltenbrunner}, {Khokhriakova}, {Kink}, {Kirsch}, {Kluge}, {Knies}, {Krippendorf}, {Krumpe}, {Kurpas}, {Li}, {Liu}, {Locatelli}, {Lorenz}, {M{\"u}ller},
  {Magaudda}, {Mannes}, {McCall}, {Meidinger}, {Michailidis}, {Migkas}, {Mu{\~n}oz-Giraldo}, {Musiimenta}, {Nguyen-Dang}, {Ni}, {Olechowska}, {Ota}, {Pacaud}, {Pasini}, {Perinati}, {Pires}, {Pommranz}, {Ponti}, {Poppenhaeger}, {P{\"u}hlhofer}, {Rau}, {Reh}, {Reiprich}, {Roster}, {Saeedi}, {Santangelo}, {Sasaki}, {Schmitt}, {Schneider}, {Schrabback}, {Schuster}, {Schwope}, {Seppi}, {Serim}, {Shreeram}, {Sokolova-Lapa}, {Starck}, {Stelzer}, {Stierhof}, {Suleimanov}, {Tenzer}, {Traulsen}, {Tr{\"u}mper}, {Tsuge}, {Urrutia}, {Veronica}, {Waddell}, {Willer}, {Wolf}, {Yeung}, {Zainab}, {Zangrandi}, {Zhang}, {Zhang}, \& {Zheng}}]{2024Merloni}
{Merloni}, A., {Lamer}, G., {Liu}, T., {et~al.} 2024, \href{http://dx.doi.org/10.1051/0004-6361/202347165}{\JournalTitle{\aap}, 682, A34}

\bibitem[{{Muzzin} {et~al.}(2013){Muzzin}, {Wilson}, {Demarco}, {Lidman}, {Nantais}, {Hoekstra}, {Yee}, \& {Rettura}}]{2013Muzzin}
{Muzzin}, A., {Wilson}, G., {Demarco}, R., {et~al.} 2013, \href{http://dx.doi.org/10.1088/0004-637X/767/1/39}{\JournalTitle{\apj}, 767, 39}

\bibitem[{{Muzzin} {et~al.}(2012){Muzzin}, {Wilson}, {Yee}, {Gilbank}, {Hoekstra}, {Demarco}, {Balogh}, {van Dokkum}, {Franx}, {Ellingson}, {Hicks}, {Nantais}, {Noble}, {Lacy}, {Lidman}, {Rettura}, {Surace}, \& {Webb}}]{2012Muzzin}
{Muzzin}, A., {Wilson}, G., {Yee}, H.~K.~C., {et~al.} 2012, \href{http://dx.doi.org/10.1088/0004-637X/746/2/188}{\JournalTitle{\apj}, 746, 188}

\bibitem[{Newville {et~al.}(2014)Newville, Stensitzki, Allen, \& Ingargiola}]{newville_matthew_2014_11813}
Newville, M., Stensitzki, T., Allen, D.~B., \& Ingargiola, A. 2014, {LMFIT: Non-Linear Least-Square Minimization and Curve-Fitting for Python}

\bibitem[{{Newville} {et~al.}(2023){Newville}, {Otten}, {Nelson}, {Stensitzki}, {Ingargiola}, {Allan}, {Fox}, {Carter}, {Micha{\l}}, {Osborn}, {Pustakhod}, {lneuhaus}, {Weigand}, {Aristov}, {Glenn}, {Deil}, {mgunyho}, {Mark}, {Hansen}, {Pasquevich}, {Foks}, {Zobrist}, {Frost}, {Stuermer}, {azelcer}, {Polloreno}, {Persaud}, {Hedegaard Nielsen}, {Pompili}, \& {Caldwell}}]{2023Newville}
{Newville}, M., {Otten}, R., {Nelson}, A., {et~al.} 2023, {lmfit/lmfit-py: 1.2.0}, Zenodo

\bibitem[{{Noble} {et~al.}(2017){Noble}, {McDonald}, {Muzzin}, {Nantais}, {Rudnick}, {van Kampen}, {Webb}, {Wilson}, {Yee}, {Boone}, {Cooper}, {DeGroot}, {Delahaye}, {Demarco}, {Foltz}, {Hayden}, {Lidman}, {Manilla-Robles}, \& {Perlmutter}}]{2017Noble}
{Noble}, A.~G., {McDonald}, M., {Muzzin}, A., {et~al.} 2017, \href{http://dx.doi.org/10.3847/2041-8213/aa77f3}{\JournalTitle{\apjl}, 842, L21}

\bibitem[{{Oguri}(2014)}]{2014Oguri}
{Oguri}, M. 2014, \href{http://dx.doi.org/10.1093/mnras/stu1446}{\JournalTitle{\mnras}, 444, 147}

\bibitem[{{Oguri} {et~al.}(2018){Oguri}, {Lin}, {Lin}, {Nishizawa}, {More}, {More}, {Hsieh}, {Medezinski}, {Miyatake}, {Jian}, {Lin}, {Takada}, {Okabe}, {Speagle}, {Coupon}, {Leauthaud}, {Lupton}, {Miyazaki}, {Price}, {Tanaka}, {Chiu}, {Komiyama}, {Okura}, {Tanaka}, \& {Usuda}}]{2018Oguri}
{Oguri}, M., {Lin}, Y.-T., {Lin}, S.-C., {et~al.} 2018, \href{http://dx.doi.org/10.1093/pasj/psx042}{\JournalTitle{\pasj}, 70, S20}

\bibitem[{Oriol {et~al.}(2023)Oriol, Virgile, Colin, Larry, J., Maxim, Ravin, Jupeng, C., A., Michael, Ricardo, Thomas, \& Robert}]{pymc2023}
Oriol, A.-P., Virgile, A., Colin, C., {et~al.} 2023, \href{http://dx.doi.org/10.7717/peerj-cs.1516}{\JournalTitle{{PeerJ} Computer Science}, 9, e1516}

\bibitem[{pandas~development team(2023)}]{the_pandas_development_team_2023_8092754}
pandas~development team, T. 2023, pandas-dev/pandas: Pandas, if you use this software, please cite it as below.

\bibitem[{{Papovich} {et~al.}(2010){Papovich}, {Momcheva}, {Willmer}, {Finkelstein}, {Finkelstein}, {Tran}, {Brodwin}, {Dunlop}, {Farrah}, {Khan}, {Lotz}, {McCarthy}, {McLure}, {Rieke}, {Rudnick}, {Sivanandam}, {Pacaud}, \& {Pierre}}]{2010Papovich}
{Papovich}, C., {Momcheva}, I., {Willmer}, C.~N.~A., {et~al.} 2010, \href{http://dx.doi.org/10.1088/0004-637X/716/2/1503}{\JournalTitle{\apj}, 716, 1503}

\bibitem[{{Persson} {et~al.}(2013){Persson}, {Murphy}, {Smee}, {Birk}, {Monson}, {Uomoto}, {Koch}, {Shectman}, {Barkhouser}, {Orndorff}, {Hammond}, {Harding}, {Scharfstein}, {Kelson}, {Marshall}, \& {McCarthy}}]{2013Persson}
{Persson}, S.~E., {Murphy}, D.~C., {Smee}, S., {et~al.} 2013, \href{http://dx.doi.org/10.1086/671164}{\JournalTitle{\pasp}, 125, 654}

\bibitem[{{Pierre} {et~al.}(2012){Pierre}, {Clerc}, {Maughan}, {Pacaud}, {Papovich}, \& {Willmer}}]{2012Pierre}
{Pierre}, M., {Clerc}, N., {Maughan}, B., {et~al.} 2012, \href{http://dx.doi.org/10.1051/0004-6361/201118169}{\JournalTitle{\aap}, 540, A4}

\bibitem[{{Planck Collaboration} {et~al.}(2016){Planck Collaboration}, {Ade}, {Aghanim}, {Arnaud}, {Ashdown}, {Aumont}, {Baccigalupi}, {Banday}, {Barreiro}, {Barrena}, {Bartlett}, {Bartolo}, {Battaner}, {Battye}, {Benabed}, {Beno{\^\i}t}, {Benoit-L{\'e}vy}, {Bernard}, {Bersanelli}, {Bielewicz}, {Bikmaev}, {B{\"o}hringer}, {Bonaldi}, {Bonavera}, {Bond}, {Borrill}, {Bouchet}, {Bucher}, {Burenin}, {Burigana}, {Butler}, {Calabrese}, {Cardoso}, {Carvalho}, {Catalano}, {Challinor}, {Chamballu}, {Chary}, {Chiang}, {Chon}, {Christensen}, {Clements}, {Colombi}, {Colombo}, {Combet}, {Comis}, {Couchot}, {Coulais}, {Crill}, {Curto}, {Cuttaia}, {Dahle}, {Danese}, {Davies}, {Davis}, {de Bernardis}, {de Rosa}, {de Zotti}, {Delabrouille}, {D{\'e}sert}, {Dickinson}, {Diego}, {Dolag}, {Dole}, {Donzelli}, {Dor{\'e}}, {Douspis}, {Ducout}, {Dupac}, {Efstathiou}, {Eisenhardt}, {Elsner}, {En{\ss}lin}, {Eriksen}, {Falgarone}, {Fergusson}, {Feroz}, {Ferragamo}, {Finelli}, {Forni}, {Frailis}, {Fraisse}, {Franceschi}, {Frejsel},
  {Galeotta}, {Galli}, {Ganga}, {G{\'e}nova-Santos}, {Giard}, {Giraud-H{\'e}raud}, {Gjerl{\o}w}, {Gonz{\'a}lez-Nuevo}, {G{\'o}rski}, {Grainge}, {Gratton}, {Gregorio}, {Gruppuso}, {Gudmundsson}, {Hansen}, {Hanson}, {Harrison}, {Hempel}, {Henrot-Versill{\'e}}, {Hern{\'a}ndez-Monteagudo}, {Herranz}, {Hildebrandt}, {Hivon}, {Hobson}, {Holmes}, {Hornstrup}, {Hovest}, {Huffenberger}, {Hurier}, {Jaffe}, {Jaffe}, {Jin}, {Jones}, {Juvela}, {Keih{\"a}nen}, {Keskitalo}, {Khamitov}, {Kisner}, {Kneissl}, {Knoche}, {Kunz}, {Kurki-Suonio}, {Lagache}, {Lamarre}, {Lasenby}, {Lattanzi}, {Lawrence}, {Leonardi}, {Lesgourgues}, {Levrier}, {Liguori}, {Lilje}, {Linden-V{\o}rnle}, {L{\'o}pez-Caniego}, {Lubin}, {Mac{\'\i}as-P{\'e}rez}, {Maggio}, {Maino}, {Mak}, {Mandolesi}, {Mangilli}, {Martin}, {Mart{\'\i}nez-Gonz{\'a}lez}, {Masi}, {Matarrese}, {Mazzotta}, {McGehee}, {Mei}, {Melchiorri}, {Melin}, {Mendes}, {Mennella}, {Migliaccio}, {Mitra}, {Miville-Desch{\^e}nes}, {Moneti}, {Montier}, {Morgante}, {Mortlock}, {Moss}, {Munshi},
  {Murphy}, {Naselsky}, {Nastasi}, {Nati}, {Natoli}, {Netterfield}, {N{\o}rgaard-Nielsen}, {Noviello}, {Novikov}, {Novikov}, {Olamaie}, {Oxborrow}, {Paci}, {Pagano}, {Pajot}, {Paoletti}, {Pasian}, {Patanchon}, {Pearson}, {Perdereau}, {Perotto}, {Perrott}, {Perrotta}, {Pettorino}, {Piacentini}, {Piat}, {Pierpaoli}, {Pietrobon}, {Plaszczynski}, {Pointecouteau}, {Polenta}, {Pratt}, {Pr{\'e}zeau}, {Prunet}, {Puget}, {Rachen}, {Reach}, {Rebolo}, {Reinecke}, {Remazeilles}, {Renault}, {Renzi}, {Ristorcelli}, {Rocha}, {Rosset}, {Rossetti}, {Roudier}, {Rozo}, {Rubi{\~n}o-Mart{\'\i}n}, {Rumsey}, {Rusholme}, {Rykoff}, {Sandri}, {Santos}, {Saunders}, {Savelainen}, {Savini}, {Schammel}, {Scott}, {Seiffert}, {Shellard}, {Shimwell}, {Spencer}, {Stanford}, {Stern}, {Stolyarov}, {Stompor}, {Streblyanska}, {Sudiwala}, {Sunyaev}, {Sutton}, {Suur-Uski}, {Sygnet}, {Tauber}, {Terenzi}, {Toffolatti}, {Tomasi}, {Tramonte}, {Tristram}, {Tucci}, {Tuovinen}, {Umana}, {Valenziano}, {Valiviita}, {Van Tent}, {Vielva}, {Villa}, {Wade},
  {Wandelt}, {Wehus}, {White}, {Wright}, {Yvon}, {Zacchei}, \& {Zonca}}]{2016Planck}
{Planck Collaboration}, {Ade}, P.~A.~R., {Aghanim}, N., {et~al.} 2016, \href{http://dx.doi.org/10.1051/0004-6361/201525823}{\JournalTitle{\aap}, 594, A27}

\bibitem[{{Planck Collaboration} {et~al.}(2020){Planck Collaboration}, {Aghanim}, {Akrami}, {Ashdown}, {Aumont}, {Baccigalupi}, {Ballardini}, {Banday}, {Barreiro}, {Bartolo}, {Basak}, {Battye}, {Benabed}, {Bernard}, {Bersanelli}, {Bielewicz}, {Bock}, {Bond}, {Borrill}, {Bouchet}, {Boulanger}, {Bucher}, {Burigana}, {Butler}, {Calabrese}, {Cardoso}, {Carron}, {Challinor}, {Chiang}, {Chluba}, {Colombo}, {Combet}, {Contreras}, {Crill}, {Cuttaia}, {de Bernardis}, {de Zotti}, {Delabrouille}, {Delouis}, {Di Valentino}, {Diego}, {Dor{\'e}}, {Douspis}, {Ducout}, {Dupac}, {Dusini}, {Efstathiou}, {Elsner}, {En{\ss}lin}, {Eriksen}, {Fantaye}, {Farhang}, {Fergusson}, {Fernandez-Cobos}, {Finelli}, {Forastieri}, {Frailis}, {Fraisse}, {Franceschi}, {Frolov}, {Galeotta}, {Galli}, {Ganga}, {G{\'e}nova-Santos}, {Gerbino}, {Ghosh}, {Gonz{\'a}lez-Nuevo}, {G{\'o}rski}, {Gratton}, {Gruppuso}, {Gudmundsson}, {Hamann}, {Handley}, {Hansen}, {Herranz}, {Hildebrandt}, {Hivon}, {Huang}, {Jaffe}, {Jones}, {Karakci}, {Keih{\"a}nen},
  {Keskitalo}, {Kiiveri}, {Kim}, {Kisner}, {Knox}, {Krachmalnicoff}, {Kunz}, {Kurki-Suonio}, {Lagache}, {Lamarre}, {Lasenby}, {Lattanzi}, {Lawrence}, {Le Jeune}, {Lemos}, {Lesgourgues}, {Levrier}, {Lewis}, {Liguori}, {Lilje}, {Lilley}, {Lindholm}, {L{\'o}pez-Caniego}, {Lubin}, {Ma}, {Mac{\'\i}as-P{\'e}rez}, {Maggio}, {Maino}, {Mandolesi}, {Mangilli}, {Marcos-Caballero}, {Maris}, {Martin}, {Martinelli}, {Mart{\'\i}nez-Gonz{\'a}lez}, {Matarrese}, {Mauri}, {McEwen}, {Meinhold}, {Melchiorri}, {Mennella}, {Migliaccio}, {Millea}, {Mitra}, {Miville-Desch{\^e}nes}, {Molinari}, {Montier}, {Morgante}, {Moss}, {Natoli}, {N{\o}rgaard-Nielsen}, {Pagano}, {Paoletti}, {Partridge}, {Patanchon}, {Peiris}, {Perrotta}, {Pettorino}, {Piacentini}, {Polastri}, {Polenta}, {Puget}, {Rachen}, {Reinecke}, {Remazeilles}, {Renzi}, {Rocha}, {Rosset}, {Roudier}, {Rubi{\~n}o-Mart{\'\i}n}, {Ruiz-Granados}, {Salvati}, {Sandri}, {Savelainen}, {Scott}, {Shellard}, {Sirignano}, {Sirri}, {Spencer}, {Sunyaev}, {Suur-Uski}, {Tauber}, {Tavagnacco},
  {Tenti}, {Toffolatti}, {Tomasi}, {Trombetti}, {Valenziano}, {Valiviita}, {Van Tent}, {Vibert}, {Vielva}, {Villa}, {Vittorio}, {Wandelt}, {Wehus}, {White}, {White}, {Zacchei}, \& {Zonca}}]{2020Planck}
{Planck Collaboration}, {Aghanim}, N., {Akrami}, Y., {et~al.} 2020, \href{http://dx.doi.org/10.1051/0004-6361/201833910}{\JournalTitle{\aap}, 641, A6}

\bibitem[{{Polletta} {et~al.}(2007){Polletta}, {Tajer}, {Maraschi}, {Trinchieri}, {Lonsdale}, {Chiappetti}, {Andreon}, {Pierre}, {Le F{\`e}vre}, {Zamorani}, {Maccagni}, {Garcet}, {Surdej}, {Franceschini}, {Alloin}, {Shupe}, {Surace}, {Fang}, {Rowan-Robinson}, {Smith}, \& {Tresse}}]{2007Polletta}
{Polletta}, M., {Tajer}, M., {Maraschi}, L., {et~al.} 2007, \href{http://dx.doi.org/10.1086/518113}{\JournalTitle{\apj}, 663, 81}

\bibitem[{{Radovich} {et~al.}(2017){Radovich}, {Puddu}, {Bellagamba}, {Roncarelli}, {Moscardini}, {Bardelli}, {Grado}, {Getman}, {Maturi}, {Huang}, {Napolitano}, {McFarland}, {Valentijn}, \& {Bilicki}}]{2017Radovich}
{Radovich}, M., {Puddu}, E., {Bellagamba}, F., {et~al.} 2017, \href{http://dx.doi.org/10.1051/0004-6361/201629353}{\JournalTitle{\aap}, 598, A107}

\bibitem[{{Rykoff} {et~al.}(2016){Rykoff}, {Rozo}, {Hollowood}, {Bermeo-Hernandez}, {Jeltema}, {Mayers}, {Romer}, {Rooney}, {Saro}, {Vergara Cervantes}, {Wechsler}, {Wilcox}, {Abbott}, {Abdalla}, {Allam}, {Annis}, {Benoit-L{\'e}vy}, {Bernstein}, {Bertin}, {Brooks}, {Burke}, {Capozzi}, {Carnero Rosell}, {Carrasco Kind}, {Castander}, {Childress}, {Collins}, {Cunha}, {D'Andrea}, {da Costa}, {Davis}, {Desai}, {Diehl}, {Dietrich}, {Doel}, {Evrard}, {Finley}, {Flaugher}, {Fosalba}, {Frieman}, {Glazebrook}, {Goldstein}, {Gruen}, {Gruendl}, {Gutierrez}, {Hilton}, {Honscheid}, {Hoyle}, {James}, {Kay}, {Kuehn}, {Kuropatkin}, {Lahav}, {Lewis}, {Lidman}, {Lima}, {Maia}, {Mann}, {Marshall}, {Martini}, {Melchior}, {Miller}, {Miquel}, {Mohr}, {Nichol}, {Nord}, {Ogando}, {Plazas}, {Reil}, {Sahl{\'e}n}, {Sanchez}, {Santiago}, {Scarpine}, {Schubnell}, {Sevilla-Noarbe}, {Smith}, {Soares-Santos}, {Sobreira}, {Stott}, {Suchyta}, {Swanson}, {Tarle}, {Thomas}, {Tucker}, {Uddin}, {Viana}, {Vikram}, {Walker}, {Zhang}, \& {DES
  Collaboration}}]{2016Rykoff}
{Rykoff}, E.~S., {Rozo}, E., {Hollowood}, D., {et~al.} 2016, \href{http://dx.doi.org/10.3847/0067-0049/224/1/1}{\JournalTitle{\apjs}, 224, 1}

\bibitem[{{Salpeter}(1955)}]{1955Salpeter}
{Salpeter}, E.~E. 1955, \href{http://dx.doi.org/10.1086/145971}{\JournalTitle{\apj}, 121, 161}

\bibitem[{{Somerville} {et~al.}(2008){Somerville}, {Hopkins}, {Cox}, {Robertson}, \& {Hernquist}}]{2008Somerville}
{Somerville}, R.~S., {Hopkins}, P.~F., {Cox}, T.~J., {Robertson}, B.~E., \& {Hernquist}, L. 2008, \href{http://dx.doi.org/10.1111/j.1365-2966.2008.13805.x}{\JournalTitle{\mnras}, 391, 481}

\bibitem[{{Somerville} \& {Kolatt}(1999)}]{1999Somerville}
{Somerville}, R.~S., \& {Kolatt}, T.~S. 1999, \href{http://dx.doi.org/10.1046/j.1365-8711.1999.02154.x}{\JournalTitle{\mnras}, 305, 1}

\bibitem[{{Somerville} {et~al.}(2015){Somerville}, {Popping}, \& {Trager}}]{2015Somerville}
{Somerville}, R.~S., {Popping}, G., \& {Trager}, S.~C. 2015, \href{http://dx.doi.org/10.1093/mnras/stv1877}{\JournalTitle{\mnras}, 453, 4337}

\bibitem[{{Stanford} {et~al.}(1995){Stanford}, {Eisenhardt}, \& {Dickinson}}]{1995Stanford}
{Stanford}, S.~A., {Eisenhardt}, P.~R.~M., \& {Dickinson}, M. 1995, \href{http://dx.doi.org/10.1086/176162}{\JournalTitle{\apj}, 450, 512}

\bibitem[{{Stanford} {et~al.}(2012){Stanford}, {Brodwin}, {Gonzalez}, {Zeimann}, {Stern}, {Dey}, {Eisenhardt}, {Snyder}, \& {Mancone}}]{2012Stanford}
{Stanford}, S.~A., {Brodwin}, M., {Gonzalez}, A.~H., {et~al.} 2012, \href{http://dx.doi.org/10.1088/0004-637X/753/2/164}{\JournalTitle{\apj}, 753, 164}

\bibitem[{{Thongkham} {et~al.}(2024){Thongkham}, {Gonzalez}, {Brodwin}, {Trudeau}, {Saha}, {Eisenhardt}, {Stanford}, {Moravec}, {Connor}, \& {Stern}}]{2024Thongkham}
{Thongkham}, K., {Gonzalez}, A.~H., {Brodwin}, M., {et~al.} 2024, \href{http://dx.doi.org/10.3847/1538-4357/ad3c44}{\JournalTitle{\apj}, 967, 123}

\bibitem[{{Virtanen} {et~al.}(2020){Virtanen}, {Gommers}, {Oliphant}, {Haberland}, {Reddy}, {Cournapeau}, {Burovski}, {Peterson}, {Weckesser}, {Bright}, {van der Walt}, {Brett}, {Wilson}, {Millman}, {Mayorov}, {Nelson}, {Jones}, {Kern}, {Larson}, {Carey}, {Polat}, {Feng}, {Moore}, {VanderPlas}, {Laxalde}, {Perktold}, {Cimrman}, {Henriksen}, {Quintero}, {Harris}, {Archibald}, {Ribeiro}, {Pedregosa}, {van Mulbregt}, \& {SciPy 1. 0 Contributors}}]{2020Virtanen}
{Virtanen}, P., {Gommers}, R., {Oliphant}, T.~E., {et~al.} 2020, \href{http://dx.doi.org/10.1038/s41592-019-0686-2}{\JournalTitle{Nature Methods}, 17, 261}

\bibitem[{{Wen} \& {Han}(2015)}]{2015Wen}
{Wen}, Z.~L., \& {Han}, J.~L. 2015, \href{http://dx.doi.org/10.1088/0004-637X/807/2/178}{\JournalTitle{\apj}, 807, 178}

\bibitem[{{Wen} \& {Han}(2021)}]{2021Wen}
{Wen}, Z.~L., \& {Han}, J.~L. 2021, \href{http://dx.doi.org/10.1093/mnras/staa3308}{\JournalTitle{\mnras}, 500, 1003}

\bibitem[{{Wen} \& {Han}(2022)}]{2022Wen}
{Wen}, Z.~L., \& {Han}, J.~L. 2022, \href{http://dx.doi.org/10.1093/mnras/stac1149}{\JournalTitle{\mnras}, 513, 3946}

\bibitem[{{Wen} \& {Han}(2024)}]{2024Wen}
{Wen}, Z.~L., \& {Han}, J.~L. 2024, \href{http://dx.doi.org/10.3847/1538-4365/ad409d}{\JournalTitle{\apjs}, 272, 39}

\bibitem[{{Wen} {et~al.}(2012){Wen}, {Han}, \& {Liu}}]{2012Wen}
{Wen}, Z.~L., {Han}, J.~L., \& {Liu}, F.~S. 2012, \href{http://dx.doi.org/10.1088/0067-0049/199/2/34}{\JournalTitle{\apjs}, 199, 34}

\bibitem[{{Werner} {et~al.}(2023){Werner}, {Cypriano}, {Gonzalez}, {Mendes de Oliveira}, {Araya-Araya}, {Doubrawa}, {Lopes de Oliveira}, {Lopes}, {Vitorelli}, {Brambila}, {Costa-Duarte}, {Telles}, {Kanaan}, {Ribeiro}, {Schoenell}, {Gon{\c{c}}alves}, {Men{\'e}ndez-Delmestre}, {Bom}, \& {Nakazono}}]{2023Werner}
{Werner}, S.~V., {Cypriano}, E.~S., {Gonzalez}, A.~H., {et~al.} 2023, \href{http://dx.doi.org/10.1093/mnras/stac3273}{\JournalTitle{\mnras}, 519, 2630}

\bibitem[{{W}es {M}c{K}inney(2010)}]{mckinney-proc-scipy-2010}
{W}es {M}c{K}inney. 2010, \href{http://dx.doi.org/10.25080/Majora-92bf1922-00a}{in {P}roceedings of the 9th {P}ython in {S}cience {C}onference, ed. {S}t\'efan van~der {W}alt \& {J}arrod {M}illman}, 56

\bibitem[{{Wilson} {et~al.}(2003){Wilson}, {Eikenberry}, {Henderson}, {Hayward}, {Carson}, {Pirger}, {Barry}, {Brandl}, {Houck}, {Fitzgerald}, \& {Stolberg}}]{2003Wilson}
{Wilson}, J.~C., {Eikenberry}, S.~S., {Henderson}, C.~P., {et~al.} 2003, \href{http://dx.doi.org/10.1117/12.460336}{in Society of Photo-Optical Instrumentation Engineers (SPIE) Conference Series, Vol. 4841, Instrument Design and Performance for Optical/Infrared Ground-based Telescopes, ed. M.~{Iye} \& A.~F.~M. {Moorwood}}, 451

\bibitem[{{Yung} {et~al.}(2023){Yung}, {Somerville}, {Finkelstein}, {Behroozi}, {Dav{\'e}}, {Ferguson}, {Gardner}, {Popping}, {Malhotra}, {Papovich}, {Rhoads}, {Bagley}, {Hirschmann}, \& {Koekemoer}}]{2023Yung}
{Yung}, L.~Y.~A., {Somerville}, R.~S., {Finkelstein}, S.~L., {et~al.} 2023, \href{http://dx.doi.org/10.1093/mnras/stac3595}{\JournalTitle{\mnras}, 519, 1578}

\end{thebibliography}
\bibliographystyle{yahapj}

\end{document}